\documentclass[11pt]{article}
\usepackage{amsfonts}
\usepackage{graphicx}
\usepackage{amsmath}
\usepackage{amssymb}
\usepackage{caption2}
\begin{document}
\bibliographystyle{prsty}
\begin{center}
{\large {\bf \sc{  Mass spectrum of the axial-vector hidden
charmed and hidden bottom tetraquark states }}} \\[2mm]
Zhi-Gang Wang \footnote{E-mail,wangzgyiti@yahoo.com.cn.  }     \\
 Department of Physics, North China Electric Power University,
Baoding 071003, P. R. China
\end{center}

\begin{abstract}
In this article, we  perform a systematic study of  the mass
spectrum of the axial-vector hidden charmed  and hidden bottom
tetraquark states using the QCD sum rules, and identify the
$Z^+(4430)$ as an axial-vector tetraquark state tentatively.
\end{abstract}



\section{Introduction}

The Babar, Belle, CLEO, D0, CDF and FOCUS collaborations have
discovered (or confirmed) a large number of charmonium-like states,
such as  $X(3940)$, $X(3872)$, $Y(4260)$, $Y(4008)$, $Y(3940)$,
$Y(4325)$, $Y(4360)$, $Y(4660)$, etc,  and revitalized  the interest
in the spectroscopy of the charmonium states
\cite{review1,review2,review3,review4,Recent-review,Olsen2009}\footnote{There
have been many theoretical works on the $X$, $Y$, $Z$ hadrons, it is
difficult to cite all of them, we prefer the comprehensive review
articles \cite{review1,review2,review3,review4,Recent-review}, where
one can find the original literatures. On the other hand, one can
consult Ref.\cite{Olsen2009} for a concise review of the
experimental situation of the new charmonium-like states.}. Many
possible assignments for those states have been suggested, such as
multiquark states (irrespective of the molecule type and the
diquark-antidiquark type), hybrid states, etc
\cite{review1,review2,review3,review4,Recent-review}.

  The $Z^+(4430)$  observed in the  decay mode $\psi^\prime\pi^+$ ($B\to\psi^\prime\pi^+K$) by the
 Belle collaboration is the most
interesting subject \cite{Belle-z4430,Belle-z4430-PRD}.  We can
distinguish the multiquark states
 from the hybrids or charmonia with the criterion of
non-zero charge. The $Z^+(4430)$ can't be a pure $c\bar{c}$ state
due to the positive charge,   and  may be a $c\bar{c}u\bar{d}$
tetraquark state. However, the Babar collaboration did not confirm
this resonance \cite{Babar0811}. Furthermore, the two resonance-like
structures  $Z(4050)$ and $Z(4250)$
 in the $\pi^+\chi_{c1}$ invariant mass distribution
near $4.1 \,\rm{GeV}$ are also particularly  interesting
\cite{Belle-chipi}. Their quark contents must be some special
combinations of the $c\bar{c} u\bar{d}$, just like the $Z^+(4430)$,
they can't be the conventional mesons.  There have been  several
theoretical interpretations for the $Z^+(4430)$, such as the
hadro-charmonium resonance \cite{review3,Voloshin0803}, the
 $S$-wave threshold effect \cite{Rosner07}, the
molecular $D^*D_1(D_1')$ state
\cite{Meng07,Lee07-sum,Liu07,Ding07,Braaten07,Liu0803,Meng0905,Ding0805},
the tetraquark state
\cite{Maiani07,Gershtein07,Li07,Liu08,Cheung0709,Bracco0807},   the
cusp in the $D^*D_1$ channel \cite{Bugg07},  the radially excited
state of the $D_s$ \cite{Matsuki0805}, the pseudo-resonance
structure \cite{Danilkin0902},  etc.

In Refs.\cite{Wang0807,Wang08072}, we assume  that the hidden
charmed mesons $Z(4050)$ and $Z(4250)$ are vector (and scalar)
tetraquark states, and study their masses using the QCD sum rules.
The numerical results indicate that the mass of the vector hidden
charmed tetraquark state is about $M_{Z}=(5.12\pm0.15)\,\rm{GeV}$ or
$(5.16\pm0.16)\,\rm{GeV}$, while the mass of the scalar hidden
charmed tetraquark state
 is about $M_{Z}=(4.36\pm0.18)\,\rm{GeV}$. The resonance-like structure  $Z(4250)$ observed by the Belle
collaboration  in the exclusive decays $\bar{B}^0\to K^- \pi^+
\chi_{c1}$  can be tentatively identified as the scalar tetraquark
state \cite{Wang08072}. In Refs.\cite{WangScalar,WangScalar-2}, we
study the mass spectrum of the scalar hidden charmed and hidden
bottom tetraquark states  in a systematic way using the QCD sum
rules. In Ref.\cite{WangVector}, we study the mass spectrum of the
vector hidden charmed and hidden bottom tetraquark states
systematically. Recently, the $0^{--}$ hidden charmed and hidden
bottom tetraquark states are studied with the QCD sum rules
\cite{Zhu2010}.

In this article, we extend our previous works to study the mass
spectrum of the axial-vector hidden charmed and hidden bottom
tetraquark states  in a systematic way with the QCD sum rules, and
make possible explanation for the nature  of the $Z^+(4430)$. The
mass is a fundamental parameter in describing a hadron, whether or
not there exist those hidden charmed or hidden bottom tetraquark
configurations is of great importance itself, because it provides a
new opportunity for a deeper understanding of the low energy QCD.
The  axial-vector hidden charmed ($c\bar{c}$) and hidden bottom
($b\bar{b}$) tetraquark states may be observed at the LHCb, where
the $b\bar{b}$ pairs will be copiously produced with the cross
section about $500 \,\mu b$ \cite{LHC}.

The hidden charmed and hidden bottom tetraquark states (denoted as
$Z$) have the symbolic quark structures:
\begin{align}
  Z^+ = Q\bar{Q} u  \bar{d}  ;~~~~
  Z^0 = \frac{1}{\sqrt{2}}Q\bar{Q}&( u  \bar{u}-d  \bar{d})  ;~~~~
  Z^- =Q\bar{Q}d\bar{u}    ; \nonumber\\
  Z_s^+ = Q\bar{Q}u  \bar{s} ;~~~~
  Z_s^- =Q\bar{Q}  s\bar{u}  ;&~~~~
  Z_s^0 = Q\bar{Q}d  \bar{s} ;~~~~
  \overline Z_s^0 = Q\bar{Q}s\bar{d} ; \nonumber \\
  Z_\varphi= \frac{1}{\sqrt{2}} Q\bar{Q} (u\bar{u}+d\bar{d});
  &~~~~Z_\phi =  Q\bar{Q} s\bar{ s} \, ,
\end{align}
where the $Q$ denote the heavy quarks $c$ and $b$.

We take the diquarks as the basic constituents   following  Jaffe
and Wilczek \cite{Jaffe2003,Jaffe2004}, and construct the
axial-vector tetraquark states with the diquark and antidiquark
pairs. The diquarks have five Dirac tensor structures, scalar
$C\gamma_5$, pseudoscalar $C$, vector $C\gamma_\mu \gamma_5$,
axial-vector $C\gamma_\mu $  and tensor $C\sigma_{\mu\nu}$, where
$C$ is the charge conjunction matrix. The structures $C\gamma_\mu $
and $C\sigma_{\mu\nu}$ are symmetric, the structures $C\gamma_5$,
$C$ and $C\gamma_\mu \gamma_5$ are antisymmetric. The attractive
interactions of one-gluon exchange favor  formation of the diquarks
in  color antitriplet $\overline{3}_{ c}$, flavor antitriplet
$\overline{3}_{ f}$ and spin singlet $1_s$ \cite{GI1,GI2}.   In this
article, we assume the axial-vector hidden charmed  and hidden
bottom tetraquark states $Z$ consist of the $C\gamma_5-C\gamma_\mu $
type rather than $C-C\gamma_\mu\gamma_5 $ type diquark structures,
and construct the interpolating currents $J^\mu(x)$ and
$\eta^\mu(x)$:
\begin{eqnarray}
J^\mu_{Z^+}(x)&=& \epsilon^{ijk}\epsilon^{imn}u_j^T(x) C\gamma_5 Q_k(x)\bar{Q}_m(x)  \gamma^\mu C \bar{d}_n^T(x)\, , \nonumber\\
J^\mu_{Z^0}(x)&=& \frac{\epsilon^{ijk}\epsilon^{imn}}{\sqrt{2}} \left[u_j^T(x) C\gamma_5 Q_k(x) \bar{Q}_m(x) \gamma^\mu C \bar{u}_n^T(x)-(u\rightarrow d)\right]\, , \nonumber\\
J^\mu_{Z^+_s}(x)&=& \epsilon^{ijk}\epsilon^{imn}u_j^T(x) C\gamma_5 Q_k(x)\bar{Q}_m(x)  \gamma^\mu C \bar{s}_n^T(x)\, , \nonumber\\
J^\mu_{Z^0_s}(x)&=& \epsilon^{ijk}\epsilon^{imn}d_j^T(x) C\gamma_5 Q_k(x)\bar{Q}_m(x)  \gamma^\mu C \bar{s}_n^T(x)\, , \nonumber\\
J^\mu_{Z_\varphi}(x)&=&\frac{\epsilon^{ijk}\epsilon^{imn}}{\sqrt{2}} \left[u_j^T(x) C\gamma_5 Q_k(x) \bar{Q}_m(x)  \gamma^\mu C \bar{u}_n^T(x) +(u\rightarrow d)\right]\, , \nonumber\\
 J^\mu_{Z_\phi}(x)&=& \epsilon^{ijk}\epsilon^{imn}s_j^T(x) C\gamma_5 Q_k(x)\bar{Q}_m(x)  \gamma^\mu C \bar{s}_n^T(x)\, ,
\end{eqnarray}

\begin{eqnarray}
\eta^\mu_{Z^+}(x)&=& \epsilon^{ijk}\epsilon^{imn}u_j^T(x) C\gamma^\mu Q_k(x)\bar{Q}_m(x)   \gamma_5C \bar{d}_n^T(x)\, , \nonumber\\
\eta^\mu_{Z^0}(x)&=& \frac{\epsilon^{ijk}\epsilon^{imn}}{\sqrt{2}}\left[u_j^T(x) C\gamma^\mu Q_k(x) \bar{Q}_m(x)  \gamma_5 C\bar{u}_n^T(x)-(u\rightarrow d)\right]\, , \nonumber\\
\eta^\mu_{Z^+_s}(x)&=& \epsilon^{ijk}\epsilon^{imn}u_j^T(x) C\gamma^\mu Q_k(x)\bar{Q}_m(x)  \gamma_5 C \bar{s}_n^T(x)\, , \nonumber\\
\eta^\mu_{Z^0_s}(x)&=& \epsilon^{ijk}\epsilon^{imn}d_j^T(x) C\gamma^\mu Q_k(x)\bar{Q}_m(x)  \gamma_5 C \bar{s}_n^T(x)\, , \nonumber\\
\eta^\mu_{Z_\varphi}(x)&=& \frac{\epsilon^{ijk}\epsilon^{imn}}{\sqrt{2}} \left[u_j^T(x) C\gamma^\mu Q_k(x) \bar{Q}_m(x)  \gamma_5 C \bar{u}_n^T(x) +(u\rightarrow d)\right]\, , \nonumber\\
 \eta^\mu_{Z_\phi}(x)&=& \epsilon^{ijk}\epsilon^{imn}s_j^T(x) C\gamma^\mu Q_k(x)\bar{Q}_m(x)  \gamma_5 C \bar{s}_n^T(x)\, ,
\end{eqnarray}
where the $i$, $j$, $k$, $\cdots$  are color indexes. In the isospin
limit, the interpolating currents result in three distinct
expressions for the spectral densities, which are characterized by
 the number of the $s$ quark they contain.
 The interpolating currents $J^\mu(x)$ and $\eta^\mu(x)$
 lead to the same expression for the
correlation functions $\Pi_{\mu\nu}(p)$, for example,
\begin{align}
  J^\mu_{Z^+} \sim  \eta^\mu_{Z^+}  ;~~~~
  J^\mu_{Z^0} \sim & \eta^\mu_{Z^0}  ;~~~~
  J^\mu_{Z^-} \sim  \eta^\mu_{Z^-}    ; \nonumber\\
  J^\mu_{Z_s^+} \sim \eta^\mu_{Z_s^+} ;~~~~
  J^\mu_{Z_s^-} \sim \eta^\mu_{Z_s^-}  ;&~~~~
  J^\mu_{Z_s^0} \sim \eta^\mu_{Z_s^0} ;~~~~
  J^\mu_{\bar{Z}_s^+} \sim \eta^\mu_{\bar{Z}_s^+} ; \nonumber \\
  J^\mu_{Z_\varphi}\sim \eta^\mu_{Z_\varphi};
  &~~~~J^\mu_{Z_\phi}\sim \eta^\mu_{Z_\phi} \, ,
\end{align}
where we use $\sim$ to denote the two interpolating  currents lead
to the same expression.   The special superpositions
$tJ^\mu(x)+(1-t)\eta^\mu(x)$   can't  improve the predictions
remarkably, where $t=0-1$. In this article, we take  the
interpolating currents $J^\mu(x)$  for simplicity, i.e. $t=1$.

In fact, we can take the colored diquarks as point particles and
describe them with  the scalar $S^a$, pseudoscalar $P^a$, vector
$V_\mu^a$, axial-vector $A_\mu^a $  and tensor $T_{\mu\nu}^a$
fields, respectively, where the $a$ is the color index, then
introduce the $SU(3)$ color interaction. We construct the color
singlet tetraquark currents  with the diquark fields $S^a$, $P^a$,
$V_\mu^a$,  $A_\mu^a $ and $T_{\mu\nu}^a$, parameterize the
nonpertubative effects with the new vacuum condensates $\langle
\overline{S}S\rangle$, $\langle \overline{P}P\rangle$, $\langle
\overline{V}V\rangle$, $\langle \overline{A}A\rangle$ and $\langle
\overline{T}T\rangle$ besides  the gluon condensate,  and  perform
the standard procedure of the QCD sum rules to study the tetraquark
states. The basic parameters such as the diquark masses and the new
vacuum condensates can be fitted phenomenally. The nonet scalar
mesons below $1\,\rm{GeV}$ (the $f_0(980)$ and $a_0(980)$
especially) are good candidates for the tetraquark states, from
those tetraquark candidates, we can obtain the basic parameters and
extend the new sum rules to other tetraquark states.  As there are
many works to do, we prefer another article.

The article is arranged as follows:  we derive the QCD sum rules for
  the axial-vector hidden charmed  and hidden bottom tetraquark states  $Z$  in Sect.2; in Sect.3, we present the
 numerical results and discussions; and Sect.4 is reserved for our
conclusions.

\section{QCD sum rules for  the axial-vector  tetraquark states $Z$ }
In the following, we write down  the two-point correlation functions
$\Pi_{\mu\nu}(p)$  in the QCD sum rules,
\begin{eqnarray}
\Pi_{\mu\nu}(p)&=&i\int d^4x e^{ip \cdot x} \langle
0|T\left[J_\mu(x)J_\nu^{\dagger}(0)\right]|0\rangle \, ,
\end{eqnarray}
where the  $J^\mu(x)$  denotes the interpolating currents
$J^\mu_{Z^+}(x)$, $J^\mu_{Z^0}(x)$, $J^\mu_{Z^+_s}(x)$, etc.

We can insert  a complete set of intermediate hadronic states with
the same quantum numbers as the current operators $J_\mu(x)$  into
the correlation functions  $\Pi_{\mu\nu}(p)$  to obtain the hadronic
representation \cite{SVZ79,Reinders85}. After isolating the ground
state contribution from the pole term of the $Z$, we get the
following result,
\begin{eqnarray}
\Pi_{\mu\nu}(p)&=&\frac{\lambda_{Z}^2}{M_{Z}^2-p^2}\left[-g_{\mu\nu}+\frac{p_\mu
p_\nu}{p^2} \right] +\cdots \, \, ,
\end{eqnarray}
where the pole residue (or coupling) $\lambda_Z$ is defined by
\begin{eqnarray}
\lambda_{Z} \epsilon_\mu  &=& \langle 0|J_\mu(0)|Z(p)\rangle \, ,
\end{eqnarray}
 the $\epsilon_\mu$ denotes the polarization vector.

 After performing the standard procedure of the QCD sum rules, we obtain the following  six sum rules:
\begin{eqnarray}
\lambda_{Z}^2 e^{-\frac{M_{Z}^2}{M^2}}= \int_{\Delta_{Z}}^{s^0_{Z}}
ds \rho_Z(s)e^{-\frac{s}{M^2}} \, ,
\end{eqnarray}
where the $Z$ denote the channels $c\bar{c}q\bar{q}$,
   $c\bar{c}q\bar{s}$, $c\bar{c}s\bar{s}$, $b\bar{b}q\bar{q}$,
   $b\bar{b}q\bar{s}$ and $b\bar{b}s\bar{s}$ respectively;
   the $s_Z^0$ are the corresponding continuum threshold parameters, and the $M^2$ is the Borel
  parameter. The thresholds $\Delta_{Z}$ can be sorted into three sets,  we introduce the $q\bar{q}$,
$q\bar{s}$ and $s\bar{s}$ to denote the light quark constituents in
the axial-vector tetraquark states to simplify the notation,
$\Delta_{q\bar{q}}=4m_Q^2$, $\Delta_{q\bar{s}}=(2m_Q+m_s)^2$,
$\Delta_{s\bar{s}}=4(m_Q+m_s)^2$.
   The explicit expressions of the  spectral densities $\rho_{q\bar{q}}(s)$,
$\rho_{q\bar{s}}(s)$ and $\rho_{s\bar{s}}(s)$ are presented in the
appendix,  where $\alpha_{f}=\frac{1+\sqrt{1-4m_Q^2/s}}{2}$,
$\alpha_{i}=\frac{1-\sqrt{1-4m_Q^2/s}}{2}$, $\beta_{i}=\frac{\alpha
m_Q^2}{\alpha s -m_Q^2}$,
$\widetilde{m}_Q^2=\frac{(\alpha+\beta)m_Q^2}{\alpha\beta}$,
$\widetilde{\widetilde{m}}_Q^2=\frac{m_Q^2}{\alpha(1-\alpha)}$.

 We carry out the operator
product expansion to the vacuum condensates adding up to
dimension-10. In calculation, we
 take   vacuum saturation for the high
dimension vacuum condensates, they  are always
 factorized to lower condensates with vacuum saturation in the QCD sum rules,
  factorization works well in  large $N_c$ limit. In reality, $N_c=3$, some  ambiguities may come from
the vacuum saturation assumption.

We take into account the contributions from the quark condensates,
mixed condensates, and neglect the contributions from the gluon
condensate. The gluon condensate $\langle
\frac{\alpha_sGG}{\pi}\rangle$  is of higher order in $\alpha_s$,
and its contributions   are suppressed by  very large denominators
comparing with the four quark condensate $\langle \bar{q}q\rangle^2$
(or $\langle \bar{s}s\rangle^2$) and would not play any significant
role, although the gluon  condensate $\langle
\frac{\alpha_sGG}{\pi}\rangle$ has smaller dimension of mass than
the four quark condensate $\langle \bar{q}q\rangle^2$ (or $\langle
\bar{s}s\rangle^2$). One can consult the sum rules for the light
tetraquark states \cite{Wang1,Wang2}, the heavy tetraquark state
\cite{Wang08072} and the  heavy molecular states
\cite{Wang0904,Wang0907} for example.  Furthermore, there are many
terms involving the gluon condensate for the heavy tetraquark states
and heavy molecular states in the operator product expansion (one
can consult Refs.\cite{Wang08072,Wang0904}), we neglect the gluon
condensate for simplicity.

In the special case of the $Y(4660)$ (as a $\psi'f_0(980)$ bound
state) and its pseudoscalar partner $\eta_c'f_0(980)$, the
contributions from the gluon condensate $\langle \frac{\alpha_s
GG}{\pi} \rangle $ are rather large \cite{WangZhang1,WangZhang2}. If
we take a simple replacement $\bar{s}(x)s(x)\rightarrow \langle
\bar{s}s\rangle$ and $\left[\bar{u}(x)u(x)+\bar{d}(x)d(x)
\right]\rightarrow 2\langle\bar{q}q\rangle$ in the interpolating
currents,  the standard  heavy quark currents $Q(x)\gamma_\mu Q(x)$
and $Q(x)i\gamma_5 Q(x)$ are obtained, where the gluon condensate
$\langle \frac{\alpha_s GG}{\pi} \rangle $ plays an important rule
in the QCD sum rules \cite{SVZ79}. The interpolating currents
constructed from the diquark-antidiquark pairs do not have such
feature. There are other interpretations  for the $Y(4660)$, for
example, the diquark-antidiquark type charmed baryonium
\cite{Cotugno-4}.

We also neglect the terms proportional to the $m_u$ and $m_d$, their
contributions are of minor importance due to the small values of the
$u$ and $d$ quark masses.

 Differentiating  the Eq.(8) with respect to  $\frac{1}{M^2}$, then eliminate the
 pole residues $\lambda_{Z}$, we can obtain the sum rules for
 the masses  of the $Z$,
 \begin{eqnarray}
 M_{Z}^2= \frac{\int_{\Delta_{Z}}^{s^0_{Z}} ds \frac{d}{d(-1/M^2)}
\rho_Z(s)e^{-\frac{s}{M^2}} }{\int_{\Delta_{Z}}^{s^0_{Z}} ds
\rho_Z(s)e^{-\frac{s}{M^2}}}\, .
\end{eqnarray}

\section{Numerical results and discussions}
The input parameters are taken to be the standard values $\langle
\bar{q}q \rangle=-(0.24\pm 0.01 \,\rm{GeV})^3$, $\langle \bar{s}s
\rangle=(0.8\pm 0.2 )\langle \bar{q}q \rangle$, $\langle
\bar{q}g_s\sigma Gq \rangle=m_0^2\langle \bar{q}q \rangle$, $\langle
\bar{s}g_s\sigma Gs \rangle=m_0^2\langle \bar{s}s \rangle$,
$m_0^2=(0.8 \pm 0.2)\,\rm{GeV}^2$,  $m_s=(0.14\pm0.01)\,\rm{GeV}$,
$m_c=(1.35\pm0.10)\,\rm{GeV}$ and $m_b=(4.8\pm0.1)\,\rm{GeV}$ at the
energy scale  $\mu=1\, \rm{GeV}$ \cite{SVZ79,Reinders85,Ioffe2005}.

The $Q$-quark masses appearing in the perturbative terms  are
usually taken to be the pole masses in the QCD sum rules, while the
choice of the $m_Q$ in the leading-order coefficients of the
higher-dimensional terms is arbitrary \cite{NarisonBook,Kho9801}.
The $\overline{MS}$ mass $m_c(m_c^2)$ relates with the pole mass
$\hat{m}_c$ through the relation $ m_c(m_c^2)
=\hat{m}_c\left[1+\frac{C_F \alpha_s(m_c^2)}{\pi}+\cdots\right]^{-1}
$. In this article, we take the approximation
$m_c(m_c^2)\approx\hat{m}_c$ without the $\alpha_s$ corrections for
consistency. The value listed in the Particle Data Group is
$m_c(m_c^2)=1.27^{+0.07}_{-0.11} \, \rm{GeV}$ \cite{PDG}, it is
reasonable to take
$\hat{m}_c=m_c(1\,\rm{GeV}^2)=(1.35\pm0.10)\,\rm{GeV}$. For the $b$
quark,  the $\overline{MS}$ mass
$m_b(m_b^2)=4.20^{+0.17}_{-0.07}\,\rm{GeV}$ \cite{PDG}, the
  gap between the energy scale $\mu=4.2\,\rm{GeV}$ and
 $1\,\rm{GeV}$ is rather large, the approximation $\hat{m}_b\approx m_b(m_b^2)\approx m_b(1\,\rm{GeV}^2)$ seems rather crude.
  It would be better to understand the quark masses $m_c$ and $m_b$ we
take at the energy scale $\mu^2=1\,\rm{GeV}^2$ as the effective
quark masses (or just the mass parameters).

In calculation, we  also neglect  the contributions from the
perturbative corrections.  Those perturbative corrections can be
taken into account in the leading logarithmic
 approximations through  anomalous dimension factors. After the Borel transform, the effects of those
 corrections are  to multiply each term on the operator product
 expansion side by the factor, $ \left[ \frac{\alpha_s(M^2)}{\alpha_s(\mu^2)}\right]^{2\Gamma_{J}-\Gamma_{\mathcal
 {O}_n}}  $,
 where the $\Gamma_{J}$ is the anomalous dimension of the
 interpolating current $J(x)$ and the $\Gamma_{\mathcal {O}_n}$ is the anomalous dimension of
 the local operator $\mathcal {O}_n(0)$. We carry out the operator product expansion at a special energy
scale $\mu^2=1\,\rm{GeV}^2$, and  set the factor $\left[
\frac{\alpha_s(M^2)}{\alpha_s(\mu^2)}\right]^{2\Gamma_{J}-\Gamma_{\mathcal
{O}_n}}\approx1$, such an approximation maybe result in some scale
dependence  and  weaken the prediction ability. In this article, we
study the axial-vector hidden charmed  and hidden bottom tetraquark
states systemically, the predictions are still robust as we take the
analogous criteria in those sum rules.

In the conventional QCD sum rules \cite{SVZ79,Reinders85}, there are
two criteria (pole dominance and convergence of the operator product
expansion) for choosing  the Borel parameter $M^2$ and threshold
parameter $s_0$. We impose the two criteria on the axial-vector
heavy tetraquark states to choose the Borel parameter $M^2$ and
threshold parameter $s_0$.

The contributions from the high dimension vacuum condensates  in the
operator product expansion are shown in Figs.1-2, where (and
thereafter) we  use the $\langle\bar{q}q\rangle$ to denote the quark
condensates $\langle\bar{q}q\rangle$, $\langle\bar{s}s\rangle$ and
the $\langle\bar{q}g_s \sigma Gq\rangle$ to denote the mixed
condensates $\langle\bar{q}g_s \sigma Gq\rangle$, $\langle\bar{s}g_s
\sigma Gs\rangle$. From the figures, we can see that  the
contributions from the high dimension condensates are very large and
change quickly with variation of the Borel parameter at the values
$M^2\leq 2.6 \,\rm{GeV}^2$ and $M^2\leq 7.2 \,\rm{GeV}^2$ in the
hidden charmed  and hidden bottom channels respectively, such an
unstable behavior cannot lead to stable sum rules, our numerical
results confirm this conjecture, see Fig.4.

At the values $M^2\geq 2.6\,\rm{GeV}^2 $ and $s_0\geq
22\,\rm{GeV}^2,\,23\,\rm{GeV}^2,\,23\,\rm{GeV}^2$, the contributions
from the $\langle \bar{q}q\rangle^2+\langle \bar{q}q\rangle \langle
\bar{q}g_s \sigma Gq\rangle $ term are less than
$12\%,\,5\%,\,2.5\%$ in the channels $c\bar{c}q\bar{q}$,
$c\bar{c}q\bar{s}$, $c\bar{c}s\bar{s}$ respectively; the
contributions from the vacuum condensate of the highest dimension
$\langle\bar{q}g_s \sigma Gq\rangle^2$ are less than
$2.5\%,\,2\%,\,1.5\%$ in the channels $c\bar{c}q\bar{q}$,
$c\bar{c}q\bar{s}$, $c\bar{c}s\bar{s}$  respectively; we expect the
operator product expansion is convergent in the hidden charmed
channels.

At the values $M^2\geq 7.2\,\rm{GeV}^2 $ and $s_0\geq
136\,\rm{GeV}^2,\,138\,\rm{GeV}^2,\,138\,\rm{GeV}^2$, the
contributions from the $\langle \bar{q}q\rangle^2+\langle
\bar{q}q\rangle \langle \bar{q}g_s \sigma Gq\rangle $ term are less
than $10\%,\,4.5\%,\,7\%$ in the channels $b\bar{b}q\bar{q}$,
$b\bar{b}q\bar{s}$, $b\bar{b}s\bar{s}$ respectively; the
contributions from the vacuum condensate of the highest dimension
$\langle\bar{q}g_s \sigma Gq\rangle^2$ are less than
$5.5\%,\,4\%,\,3\%$ in the channels $b\bar{b}q\bar{q}$,
$b\bar{b}q\bar{s}$, $b\bar{b}s\bar{s}$  respectively; we expect the
operator product expansion is convergent in the hidden bottom
channels.

 In this article, we take the uniform Borel parameter
$M^2_{min}$, i.e. $M^2_{min}\geq 2.6 \, \rm{GeV}^2$  and
$M^2_{min}\geq 7.2 \, \rm{GeV}^2$ in the hidden charmed   and hidden
bottom channels respectively.

In Fig.3, we show the  contributions from the pole terms with
variation of the Borel parameters and the threshold parameters. The
pole contributions are larger than (or equal) $46\%,\,50\%,\,50\%$
at the value $M^2 \leq 3.2 \, \rm{GeV}^2 $ and $s_0\geq
22\,\rm{GeV}^2,\,23\,\rm{GeV}^2,\,23\,\rm{GeV}^2 $
 in the channels $c\bar{c}q\bar{q}$, $c\bar{c}q\bar{s}$,  $c\bar{c}s\bar{s}$
 respectively, and larger than (or equal)
$48\%,\,50\%,\,50\%$ at the value $M^2 \leq 8.2 \, \rm{GeV}^2 $ and
$s_0\geq 136\,\rm{GeV}^2,\,138\,\rm{GeV}^2,\,138\,\rm{GeV}^2 $
 in  the channels $b\bar{b}q\bar{q}$,  $b\bar{b}q\bar{s}$,
$b\bar{b}s\bar{s}$  respectively. Again we take the uniform Borel
parameter $M^2_{max}$, i.e. $M^2_{max}\leq 3.2 \, \rm{GeV}^2$ and
$M^2_{max}\leq 8.2 \, \rm{GeV}^2$ in the hidden charmed  and hidden
bottom channels respectively.

In this article, the threshold parameters are taken as
$s_0=(23\pm1)\,\rm{GeV}^2$, $(24\pm1)\,\rm{GeV}^2$,
$(24\pm1)\,\rm{GeV}^2$, $(138\pm2)\,\rm{GeV}^2$,
$(140\pm2)\,\rm{GeV}^2$, $(140\pm2)\,\rm{GeV}^2$ in the channels
$c\bar{c}q\bar{q}$, $c\bar{c}q\bar{s}$,
    $c\bar{c}s\bar{s}$, $b\bar{b}q\bar{q}$, $b\bar{b}q\bar{s}$,
      $b\bar{b}s\bar{s}$  respectively;
   the Borel parameters are taken as $M^2=(2.6-3.2)\,\rm{GeV}^2$ and
   $(7.2-8.2)\,\rm{GeV}^2$ in the
hidden charmed and hidden bottom channels respectively.
      In those regions,  the pole contributions are about  $(46-74)\%$,
$(50-77)\%$, $(50-77)\%$,  $(48-67)\%$, $(50-69)\%$,  $(50-69)\%$ in
the channels $c\bar{c}q\bar{q}$, $c\bar{c}q\bar{s}$,
    $c\bar{c}s\bar{s}$, $b\bar{b}q\bar{q}$, $b\bar{b}q\bar{s}$,
      $b\bar{b}s\bar{s}$  respectively;   the two criteria of the QCD sum rules
are fully  satisfied  \cite{SVZ79,Reinders85}.

From Fig.3, we can see that the Borel windows $M_{max}^2-M_{min}^2$
change with variations of the  threshold parameters $s_0$. In this
article, the Borel windows  are taken as $0.6\,\rm{GeV}^2$ and
$1.0\,\rm{GeV}^2$ in the hidden charmed  and hidden bottom channels
respectively; they are small enough.  If we take larger threshold
parameters,  the Borel windows are larger and the resulting  masses
are larger, see Fig.4. In this article, we intend to  calculate the
possibly  lowest masses which are supposed to be the ground state
masses  by imposing the two criteria of the QCD sum rules.

If we take  analogous pole contributions,  the interpolating current
with more $s$ quarks requires slightly larger threshold parameter
due to the $SU(3)$ breaking effects, see Fig.3. In the channels
$Q\bar{Q}q\bar{s}$ and $Q\bar{Q}s\bar{s}$, the $SU(3)$ breaking
effects on the threshold parameters are tiny, we take uniform
threshold parameters in those channels. In Fig.4, we plot the
axial-vector tetraquark state masses $M_Z$ with variation of the
Borel parameters and the threshold parameters.  Naively, we expect
the tetraquark state with more $s$ quarks will have larger mass. In
calculations, we observe that the possibly  lowest masses of the
axial-vector heavy tetraquark states $Q\bar{Q}q\bar{s}$ and
$Q\bar{Q}s\bar{s}$ are almost the same.

Taking into account all uncertainties of the relevant  parameters,
finally we obtain the values of the masses and pole resides of
 the axial-vector tetraquark states  $Z$, which are  shown in Figs.5-6 and Table 1.
In Table 1, we also present the masses of the scalar hidden charmed
and hidden bottom tetraquark states obtained in our previous works
\cite{WangScalar,WangScalar-2}.

In this article,  we calculate the uncertainties $\delta$  with the
formula
\begin{eqnarray}
\delta=\sqrt{\sum_i\left(\frac{\partial f}{\partial
x_i}\right)^2\mid_{x_i=\bar{x}_i} (x_i-\bar{x}_i)^2}\,  ,
\end{eqnarray}
 where the $f$ denote  the
hadron mass  $M_Z$ and the pole residue $\lambda_Z$,  the $x_i$
denote the relevant parameters $m_c$, $m_b$, $\langle \bar{q}q
\rangle$, $\langle \bar{s}s \rangle$, $\cdots$. As the partial
 derivatives   $\frac{\partial f}{\partial x_i}$ are difficult to carry
out analytically, we take the  approximation $\left(\frac{\partial
f}{\partial x_i}\right)^2 (x_i-\bar{x}_i)^2\approx
\left[f(\bar{x}_i\pm \Delta x_i)-f(\bar{x}_i)\right]^2$ in the
numerical calculations.

From Table 1, we can see that the uncertainties of the masses $M_Z$
are rather small (about $4\%$ in the hidden charmed  channels and
$2\%$ in the hidden bottom channels),  while the uncertainties of
the pole residues $\lambda_{Z}$ are rather large (about $20\%$). The
uncertainties of the input parameters ($\langle \bar{q}q \rangle$,
$\langle \bar{s}s \rangle$, $\langle \bar{s}g_s\sigma G s \rangle$,
$\langle \bar{q}g_s\sigma G q \rangle$,  $m_s$,  $m_c$ and $m_b$)
vary in the range $(2-25)\%$, the uncertainties of the pole residues
$\lambda_{Z}$ are reasonable. We obtain the  squared masses $M_Z^2$
through a fraction,  the uncertainties in the numerator and
denominator which originate  from a given input parameter (for
example, $\langle \bar{s}s \rangle$, $\langle \bar{s}g_s\sigma G s
\rangle$) cancel out with each other, and result in small net
uncertainty.

The $SU(3)$ breaking effects for the masses of the axial-vector
 hidden charmed  and hidden bottom tetraquark states are buried in the
uncertainties. Naively, we expect the axial-vector and vector
diquarks have larger masses than the corresponding scalar diquarks,
and the masses of the tetraquark states  have the hierarchy
$M_{C\gamma_\mu-C\gamma^\mu}\geq M_{C\gamma_5-C\gamma^\mu} \geq
M_{C\gamma_5-C\gamma_5}$, because the attractive interactions of
one-gluon exchange favor formation of the diquarks in  color
antitriplet $\overline{3}_{ c}$, flavor antitriplet $\overline{3}_{
f}$ and spin singlet $1_s$ \cite{GI1,GI2}.  From Table 1, we can see
that it is not the case.

In the conventional QCD sum rules, we usually consult the
experimental data in choosing the Borel parameter $M^2$ and the
threshold parameter $s_0$. If the mass spectrum of the axial-vector
tetraquark states are well known, we can denote  the ground state,
the first excited state, the second excited state, the third excited
state, $\dots$, as $Z$, $Z'$, $Z''$, $Z'''$, $\dots$. The critical
thresholds for  emergence of those excited  tetraquark states are
$T_{Z'}$, $T_{Z''}$, $T_{Z'''}$, $\dots$, respectively. The
threshold parameter $s_0$ should take values in the region $
(M_Z+\Gamma_Z)^2\leq s_0<T_{Z'}$. However, the present experimental
knowledge about the phenomenological hadronic spectral densities of
the multiquark states is  rather vague, even the existence of the
multiquark states is not confirmed with confidence, and no knowledge
about either there are high resonances or not.

Taking into account the two criteria (pole dominance and convergence
of the operator product expansion) of the QCD sum rules, we  can
obtain the possibly lowest threshold parameter $s_0$, which is
denoted as $s^0_{min}$. In this article, we take the value
$s^0_{min}$ and make crude estimations for the ground state masses.

 The values of the
$s^0_{min}$ in different channels  maybe smaller (or larger) than
$T_{Z'}$, or even smaller than $(M_Z+\Gamma_Z)^2$, the two criteria
of the QCD sum rules alone cannot always warrant satisfactory
threshold parameters and Borel windows. For example, the nonet
scalar mesons below $1\,\rm{GeV}$ (the $f_0(980)$ and $a_0(980)$
especially) are good candidates for the tetraquark states
\cite{Jaffe2004,Close2002,ReviewScalar}. The two criteria of the QCD
sum rules result in the threshold parameters $s_0\gg
(M_{f_0/a_0}+\Gamma_{f_0/a_0})^2$, the contributions of the excited
states are already included in if there are any, and we have to
resort to "multi-pole $+$ continuum states" to approximate the
phenomenological spectral densities. If we insist on the "one-pole
$+$ continuum states"
 ansatz, no reasonable Borel window can be obtained, although it is not an indication
non-existence of the light tetraquark states (For detailed
discussions about this subject, one can consult
Refs.\cite{Wang08072,Wang0708}). The QCD sum rules is just a QCD
model.

In the channel $c\bar{c}q\bar{q}$, the  threshold parameter
 $s^0_{min}$ leads to the mass  $M_{c\bar{c}q\bar{q}}=(4.32\pm0.18)\,\rm{GeV}$, which is consistent
 with the experimental data $M_Z=(4433\pm4\pm2)\,\rm{MeV}$ or $4443^{+15}_{-12}{^{+19}_{-13}}\,\rm{MeV}$
from the Belle collaboration within uncertainty
\cite{Belle-z4430,Belle-z4430-PRD}. The experimental value is
$(M_Z+\Gamma_Z)^2\leq 22.5\,\rm{GeV}^2$, the lower bound of the
$s^0_{min}=(22-24)\,\rm{GeV}^2$ is smaller than $(M_Z+\Gamma_Z)^2$,
we have to postpone the $s^0_{min}$ to larger values. If we take
$s_0=(26\pm1)\,\rm{GeV}^2$, the prediction
$M_Z=(4.44\pm0.19)\,\rm{GeV}$ is in excellent agreement with
experimental data, see Fig.7. In Fig.8, we present the corresponding
pole residue, from the figure, we can see that larger threshold
parameter result in larger pole residue.

The predictions of the QCD sum rules favor the scenario of the
$Z^+(4430)$ as an axial-vector tetraquark state, the $Z^+(4430)$ can
be tentatively identified as an axial-vector tetraquark state. In
other channels, the heavy axial-vector tetraquark states exist in
nature maybe have larger masses than those theoretical predictions
presented in Table 1. On the other hand, the upper bound of the
threshold parameter $s_0=(25-27)\,\rm{GeV}^2$ maybe larger than the
critical threshold $T_{Z'}$,  so we identify the $Z^+(4430)$ as an
axial-vector tetraquark state tentatively, not confidently.

In Refs.\cite{Wang08072,WangScalar,WangScalar-2}, we observe that
the meson $Z(4250)$ may be a scalar tetraquark state
($c\bar{c}u\bar{d}$), irrespective of the $C\gamma_\mu-C\gamma^\mu$
type and the $C\gamma_5-C\gamma_5$ type diquark structures, the
decay $ Z(4250) \to \pi^+\chi_{c1}$ can take place with the
Okubo-Zweig-Iizuka   super-allowed "fall-apart" mechanism, which can
take into account the large total width naturally.  In the present
case, the decay $Z(4430)\to \psi'\pi$ can also take place with the
Okubo-Zweig-Iizuka  super-allowed "fall-apart" mechanism, which can
take into account the large total width
($\Gamma_Z=45^{+18}_{-13}{^{+30}_{-13}}\,\rm{MeV}$ or
$107^{+86}_{-43}{^{+74}_{-56}}\,\rm{MeV}$) naturally
\cite{Belle-z4430,Belle-z4430-PRD}.

In this article, we take the simple pole $+$ continuum approximation
for the phenomenological spectral  densities. In fact, such a simple
approximation has shortcomings. In the case of the non-relativistic
harmonic-oscillator potential model,  the spectrum of the bound
states (the masses $E_n$ and the wave functions $\Psi_n(x)$) and the
exact correlation functions (and hence its operator product
expansion to any order) are known precisely. The non-relativistic
harmonic-oscillator potential $\frac{1}{2}m\omega^2\vec{r}^2$ is
highly non-perturbative, one suppose the full Green function
satisfies the Lippmann-Schwinger operator equation and may be solved
perturbatively. We can introduce the Borel parameter dependent
effective threshold parameter
$z_{eff}(M)=\omega\left[\bar{z}_0+\bar{z}_1 \sqrt{ \frac{\omega}{M}
}+\bar{z}_2 \frac{\omega}{M}+\cdots \right]$ and fit the
coefficients $\bar{z}_i$ to reproduce both the ground energy $E_0$
and the pole residue $R_0=\Psi_0^*(0)\Psi_0(0)$,  the
phenomenological spectrum density can be described by the
perturbative contributions well above the effective continuum
threshold  $z_{eff}(M)$, or reproduce the ground energy $E_0$ only
and take the pole residue $R$ as a calculated parameter, there
exists a solution for the effective continuum threshold $z_{eff}(M)$
which precisely reproduces the exact ground energy $E_0$
 for any value of the pole residue $R$ within the range $R=(0.7-1.15)R_0$ in the
limited fiducial Borel window,   the value of the pole residue $R$
 extracted from the sum rule is
determined to a great extent by the contribution of the hadron
continuum \cite{ChSh3}. There maybe systemic uncertainties  out of
control.

In the real QCD world, the hadronic spectral densities are not known
exactly. In the present case, the ground states have not been
observed yet, except for the possible  axial-vector tetraquark state
candidate $Z^+(4430)$. So we have no confidence to introduce the
Borel parameter dependent effective threshold parameter
$s_{eff}(M)=\bar{s}_0+\bar{s}_1 \frac{1}{M^2} +\bar{s}_2
\frac{1}{M^4}+\cdots $ and approximate the phenomenological spectral
densities  with the perturbative contributions above the effective
continuum  threshold $s_{eff}(M)$ accurately. Furthermore, the pole
residues (or the couplings of the interpolating  currents to the
ground state tetraquark) $\lambda_{Z}$ are not experimentally
measurable quantities, and should be calculated by some theoretical
approaches, the true values are difficult to obtain,  which are
distinguished from the decay constants of the pseudoscalar mesons
and the vector mesons, the decay constants can be measured with
great precision in the leptonic decays (in some channels).

The  spectrum of the bound states in the non-relativistic
harmonic-oscillator potential model are of the Dirac $\delta$
function type, we can choose $z_{eff}<E_1$, while in the case of the
QCD, the situation is  rather complex, the effective continuum
thresholds $s_{eff}(M)$ maybe overlap with the first radial excited
states, which are usually broad. For example, in the pseudoscalar
channels, the widths of the  $\pi$, $\pi(1300)$, $\pi(1800)$,
$\cdots$ are $\sim 0\,\rm{GeV}$, $(0.2-0.6)\,\rm{GeV}$,
$(0.208\pm0.012)\,\rm{GeV}$, $\cdots$ respectively, while  the
widths of the $K$, $K(1460)$, $K(1830)$, $\cdots$ are $\sim 0
\,\rm{GeV}$, $\sim (0.25-0.26)\,\rm{GeV}$, $\sim 0.25\,\rm{GeV}$,
$\cdots$ respectively \cite{PDG}. In this article, we prefer (or
have to choose) the simple pole $+$ continuum approximation, and
cannot estimate the unknown systemic uncertainties of the QCD sum
rules before the spectral densities in both the QCD and
phenomenological sides are  known with great accuracy.

 In Ref.\cite{LuchaPLB}, Lucha, Melikhov and Simula
 use the correlation  function of the pseudoscalar current $J_5(x)=(m_b+m_u)\bar{q}(x)i\gamma_5b(x)$
 to illustrate  a Borel-parameter-dependent effective
continuum threshold can  reduce considerably the (unphysical)
dependence of the extracted bound-state mass and the decay constant
on the Borel parameter. In the present case, we have no experimental
data for the masses and pole residues of the tetraquark states to
take  as a guide and apply the $\chi^2$  minimization  by adjusting
the effective threshold parameters. On the other hand, the
Borel-parameter-dependent effective continuum thresholds maybe
overlap with the $(M_Z+\Gamma_Z)^2$, $T_{Z'}$, $T_{Z''}$, $\cdots$,
we prefer  the Borel-parameter-independent threshold parameters,
although the Borel-parameter-dependent effective threshold
parameters maybe smear some dependence on the Borel parameter. From
Figs.5-8, we can see that the dependence  of the masses and pole
residues on the Borel parameters in the Borel windows are rather
mild.

The central values of our predictions are much larger than the
corresponding ones from  a relativistic quark model based on a
quasipotential approach in QCD \cite{Ebert05,Ebert0808}. In
Refs.\cite{Ebert05,Ebert0808}, Ebert et al take the diquarks as
bound states of the light and heavy quarks in the color antitriplet
channel, and calculate their mass spectrum using a Schrodinger type
equation, then take the masses of the diquarks  as the basic input
parameters, and study the mass spectrum of the heavy tetraquark
states as bound states of the diquark-antidiquark system.  In
Refs.\cite{Maiani2004,Maiani20042,Polosa0902}, Maiani et al take the
diquarks as the basic constituents, examine the rich spectrum of the
 diquark-antidiquark states  from the constituent diquark masses and the spin-spin
 interactions, and try to  accommodate some of the newly observed charmonium-like resonances not
 fitting a pure $c\bar{c}$ assignment; furthermore, the corresponding bottom tetraquark states are also studied with
 the same method \cite{Ali-2010}.  The predictions depend heavily on  the assumption that the light
 scalar mesons $a_0(980)$ and $f_0(980)$ are tetraquark states,
 the  basic  parameters (constituent diquark masses) are
 estimated thereafter.
 In the conventional quark models, the
constituent quark masses  are taken as the basic input parameters,
and fitted to reproduce the mass spectra  of the conventional mesons
and baryons. However, the present experimental knowledge about the
phenomenological hadronic spectral densities of the tetraquark
states is  rather vague, whether or not there exist   tetraquark
states is not confirmed with confidence, and no knowledge about  the
high resonances. The predicted constituent diquark masses can not be
confronted with the experimental data.

In Refs.\cite{Maiani07,Ebert0808,Maiani20042}, the $X(3872)$ and
$Z(4430)$ are taken as the ground state axial-vector and first
radially  excited axial-vector tetraquark states respectively. In
Ref.\cite{Matheus0608}, Matheus et al study the $X(3872)$ as a
tetraquark state with $J^{PC}=1^{++}$ using QCD spectral sum rules,
the prediction is consistent with the experimental data within
uncertainty. The discrepancy between  the predictions of
Ref.\cite{Matheus0608} and the present work (analogous interpolating
currents are chosen in those works) mainly originates from the high
dimensional vacuum condensates $\langle \bar{q}g_s \sigma G
q\rangle^2$  which are neglected in Ref.\cite{Matheus0608}. The
condensates $\langle \bar{q}g_s \sigma G q\rangle^2$ are counted as
$\mathcal {O}(\frac{m_c^6}{M^6})$, and the corresponding
contributions are greatly enhanced  at small $M^2$, and result in
rather  bad  convergent behavior in the operator product expansion,
we have to choose larger Borel parameter $M^2$. We insist on taking
into account the high dimensional  vacuum condensates, as the
interpolating current consists  of  a light quark-antiquark pair and
a heavy quark-antiquark pair, one of the highest dimensional vacuum
condensates is $\langle \bar{q}q\rangle^2\times \langle
\frac{\alpha_s GG}{\pi}\rangle$, we have to take into account the
condensates $\langle \bar{q}g_s \sigma G q\rangle^2$ for
consistence.

\begin{figure}
 \centering
 \includegraphics[totalheight=5cm,width=6cm]{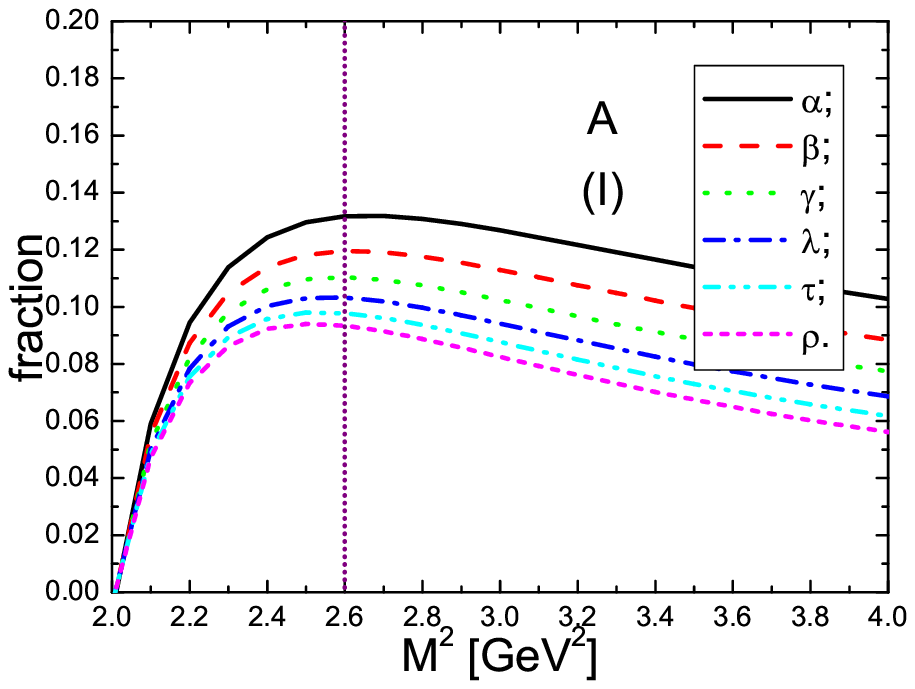}
 \includegraphics[totalheight=5cm,width=6cm]{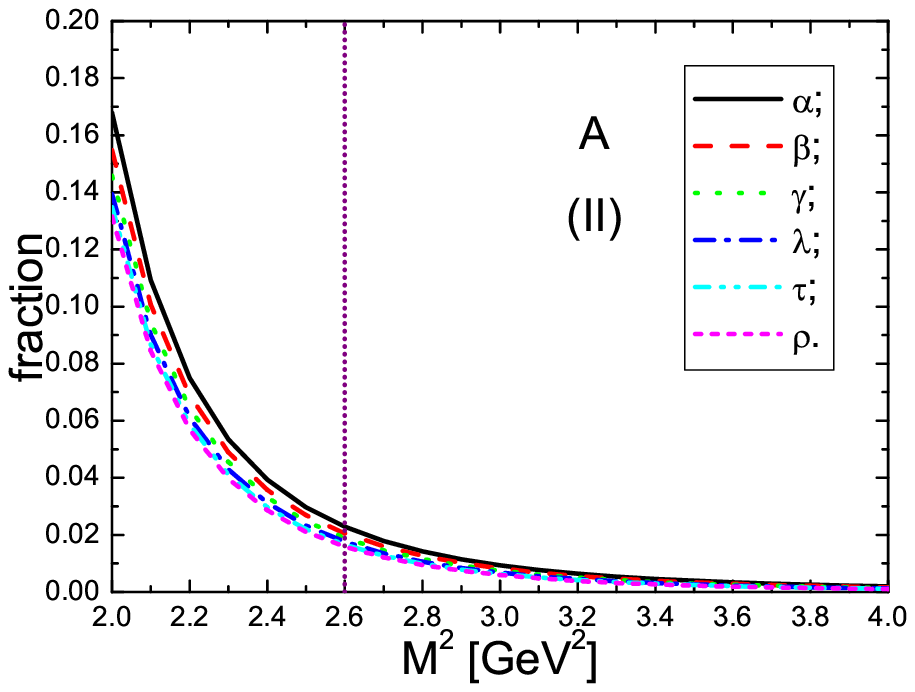}
 \includegraphics[totalheight=5cm,width=6cm]{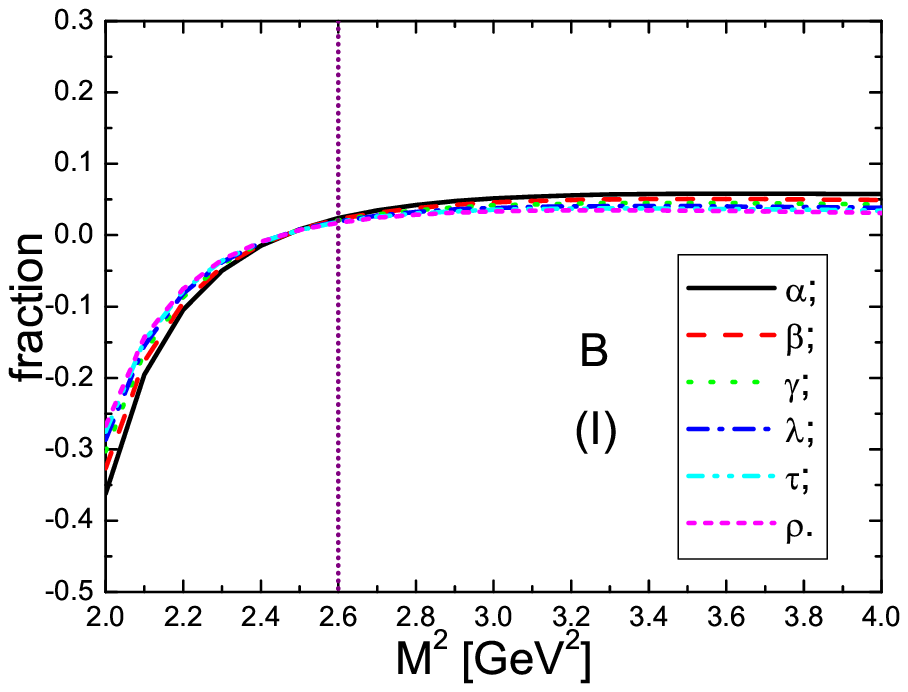}
 \includegraphics[totalheight=5cm,width=6cm]{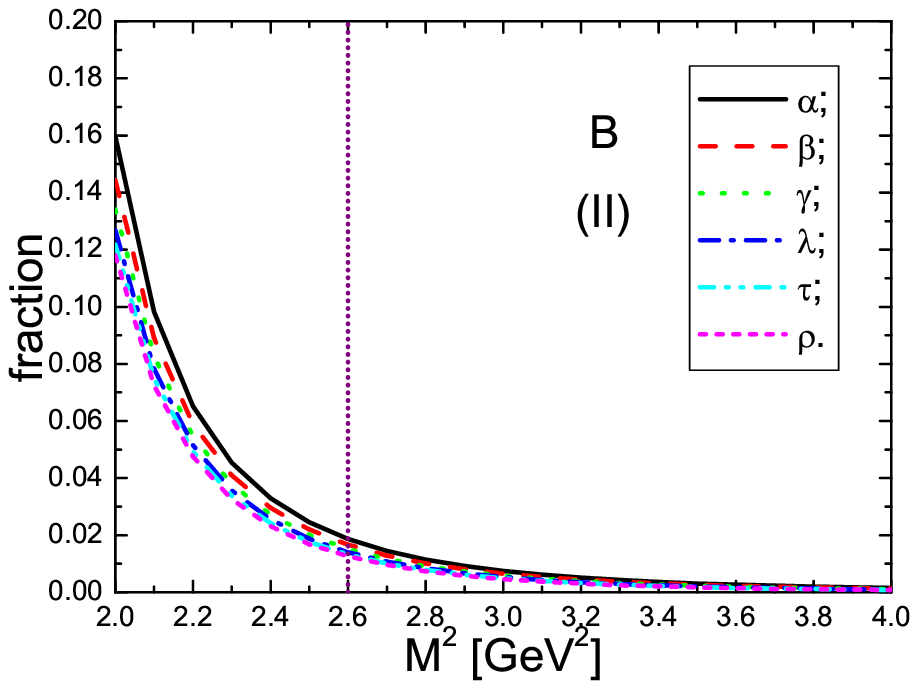}
  \includegraphics[totalheight=5cm,width=6cm]{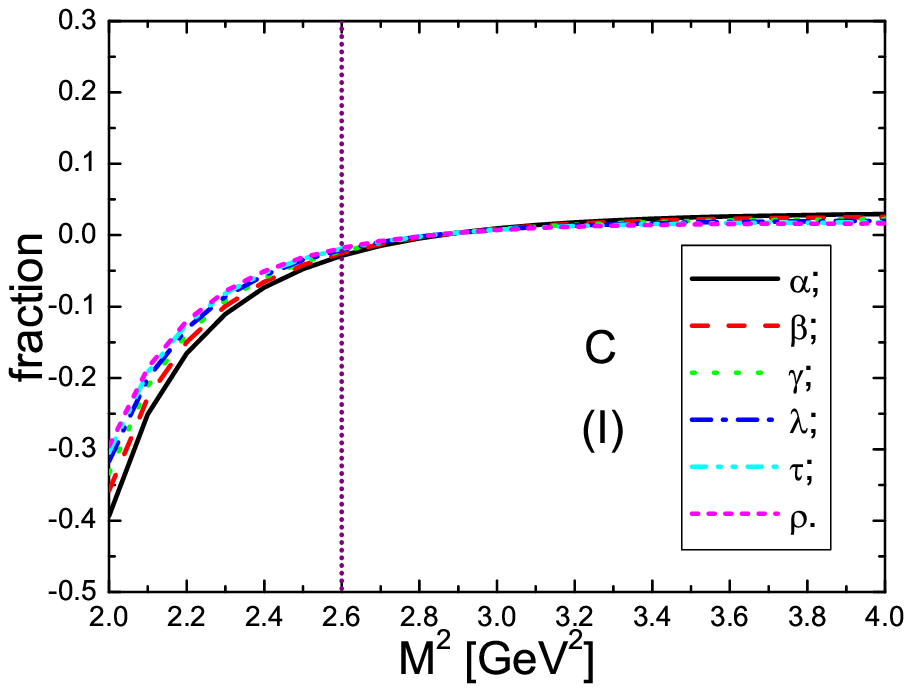}
   \includegraphics[totalheight=5cm,width=6cm]{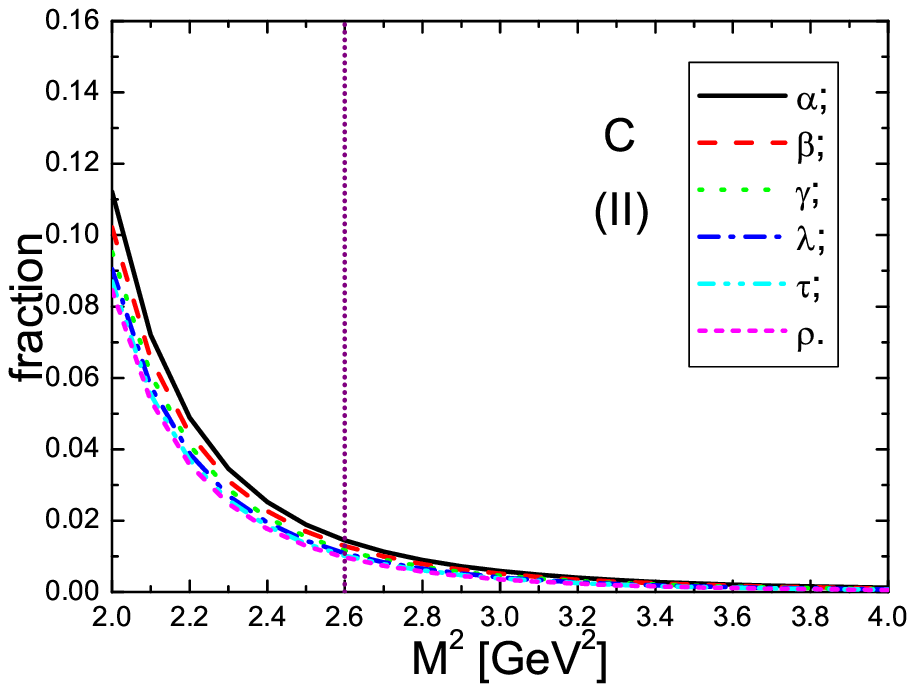}
   \caption{ The contributions from different terms with variation of the Borel
   parameter $M^2$  in the operator product expansion. The (I)  and (II) denote the contributions from the
   $\langle \bar{q}q\rangle^2+\langle \bar{q}q\rangle \langle \bar{q}g_s \sigma Gq\rangle
   $ term and the  $ \langle \bar{q}g_s \sigma Gq\rangle^2
   $ term  respectively. The $A$, $B$ and $C$
    denote the channels $c\bar{c}q\bar{q}$,
   $c\bar{c}q\bar{s}$ and $c\bar{c}s\bar{s}$  respectively. The notations
   $\alpha$, $\beta$, $\gamma$, $\lambda$, $\tau$ and $\rho$  correspond to the threshold
   parameters $s_0=21\,\rm{GeV}^2$,
   $22\,\rm{GeV}^2$, $23\,\rm{GeV}^2$, $24\,\rm{GeV}^2$, $25\,\rm{GeV}^2$ and $26\,\rm{GeV}^2$ respectively.}
\end{figure}

\begin{figure}
 \centering
 \includegraphics[totalheight=5cm,width=6cm]{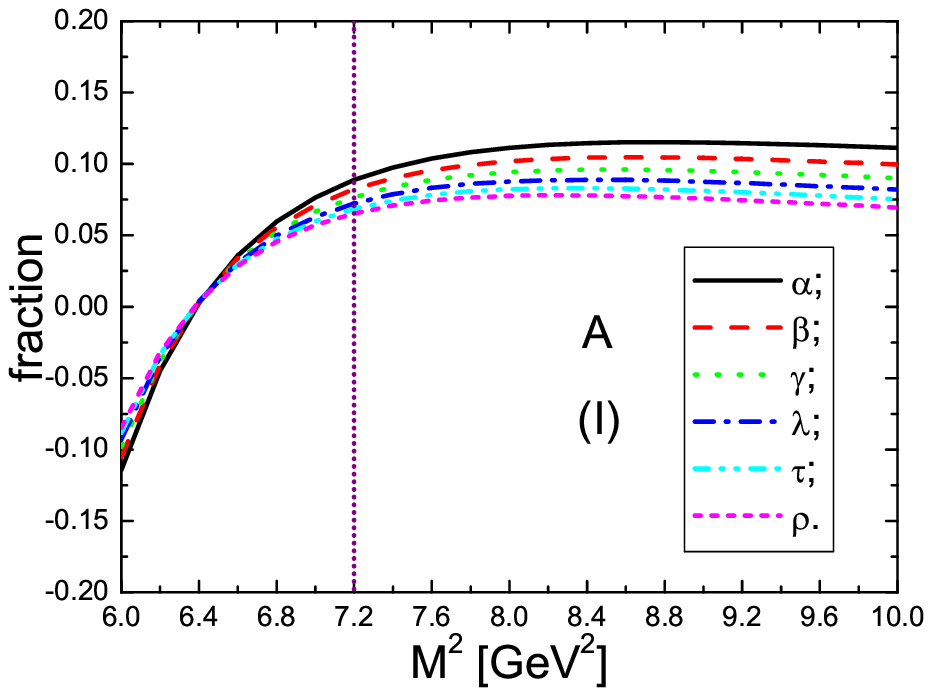}
 \includegraphics[totalheight=5cm,width=6cm]{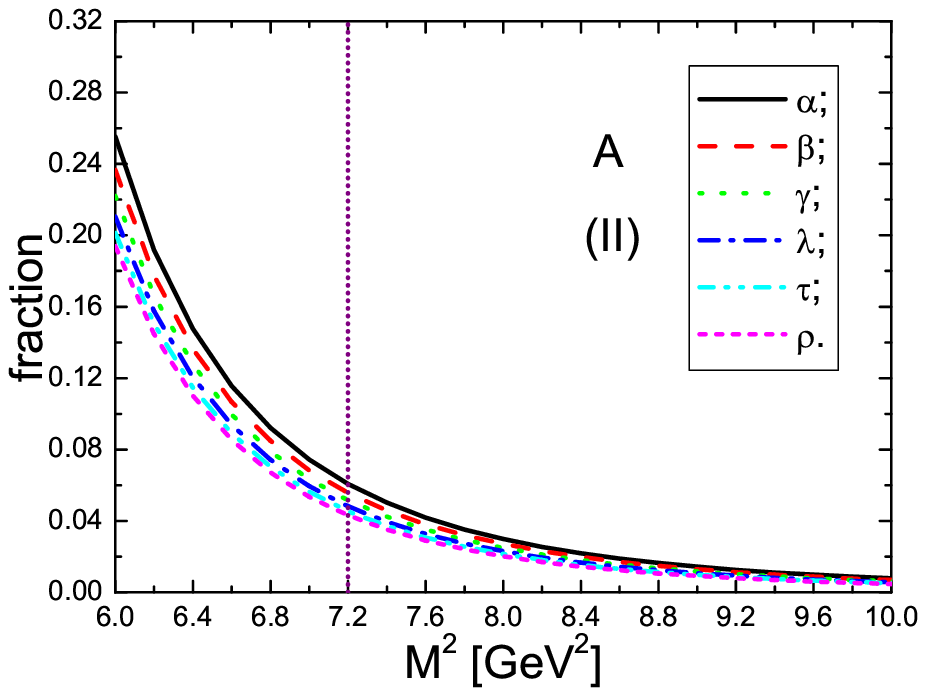}
 \includegraphics[totalheight=5cm,width=6cm]{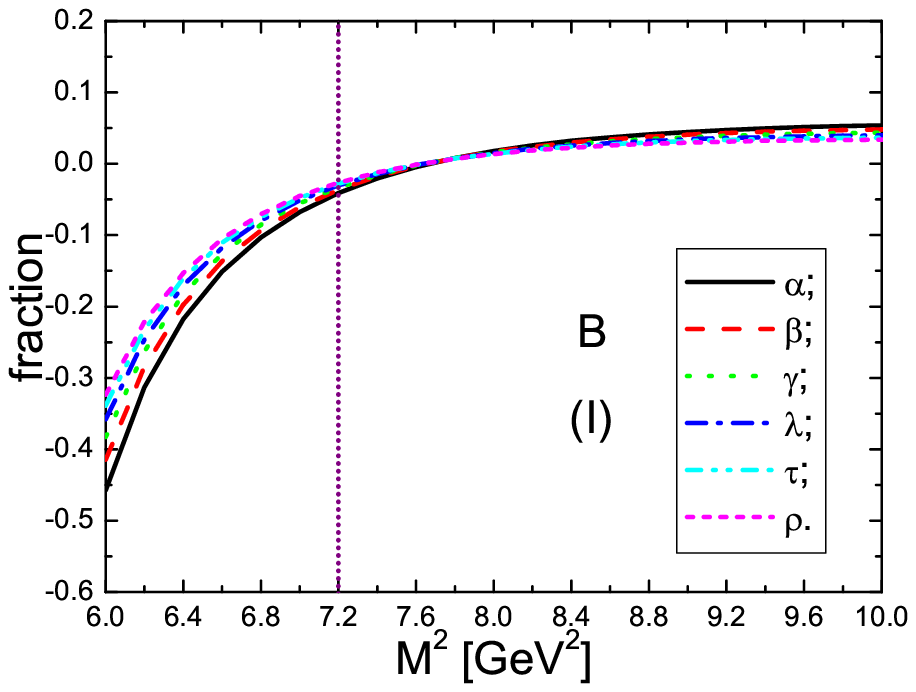}
 \includegraphics[totalheight=5cm,width=6cm]{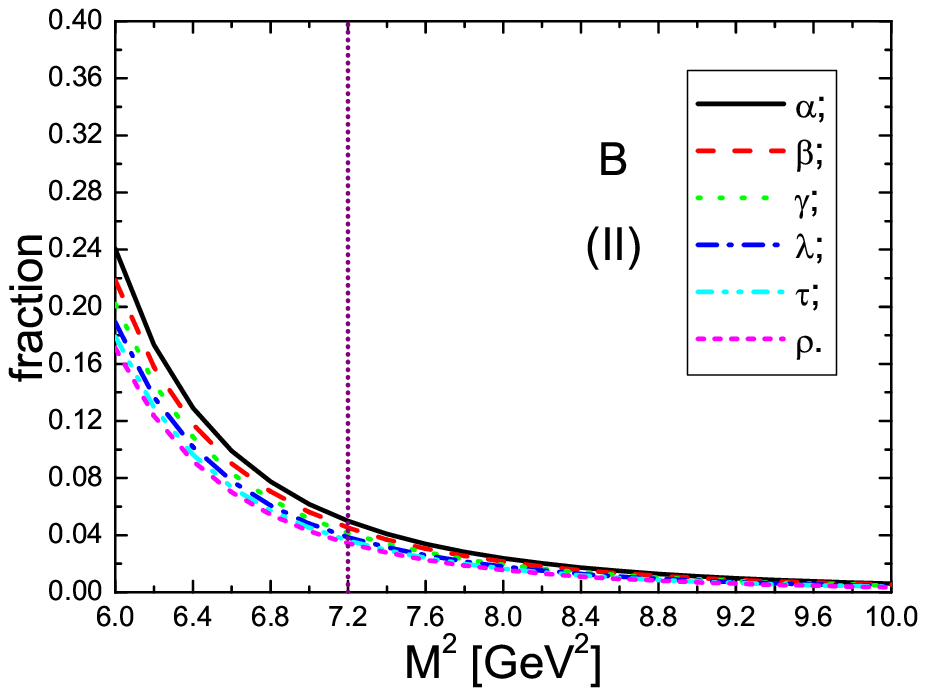}
  \includegraphics[totalheight=5cm,width=6cm]{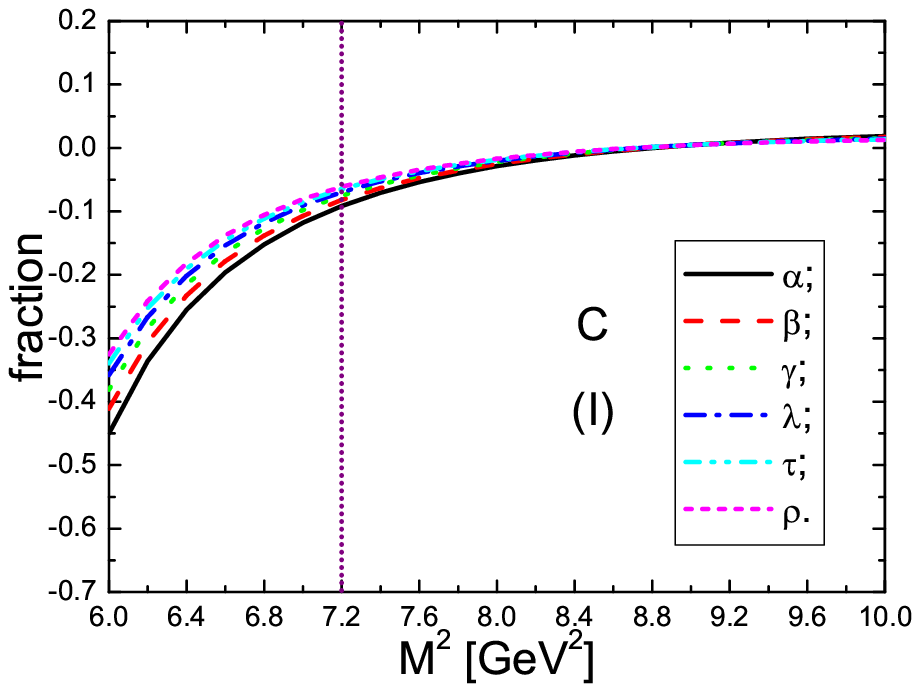}
   \includegraphics[totalheight=5cm,width=6cm]{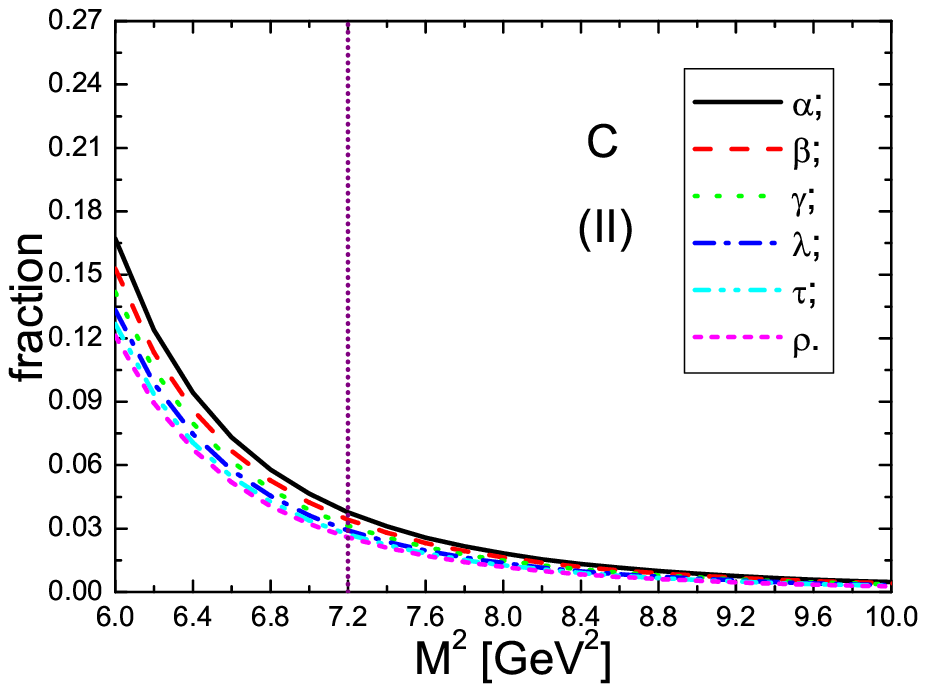}
   \caption{ The contributions from different terms with variation of the Borel
   parameter $M^2$  in the operator product expansion. The (I)  and (II) denote the contributions from the
   $\langle \bar{q}q\rangle^2+\langle \bar{q}q\rangle \langle \bar{q}g_s \sigma Gq\rangle
   $ term and the  $ \langle \bar{q}g_s \sigma Gq\rangle^2
   $ term  respectively. The $A$, $B$ and $C$
    denote the channels $b\bar{b}q\bar{q}$,
   $b\bar{b}q\bar{s}$ and $b\bar{b}s\bar{s}$  respectively. The notations
   $\alpha$, $\beta$, $\gamma$, $\lambda$, $\tau$ and $\rho$  correspond to the threshold
   parameters $s_0=132\,\rm{GeV}^2$,
   $134\,\rm{GeV}^2$, $136\,\rm{GeV}^2$, $138\,\rm{GeV}^2$, $140\,\rm{GeV}^2$ and $142\,\rm{GeV}^2$ respectively.}
\end{figure}

\begin{figure}
 \centering
 \includegraphics[totalheight=5cm,width=6cm]{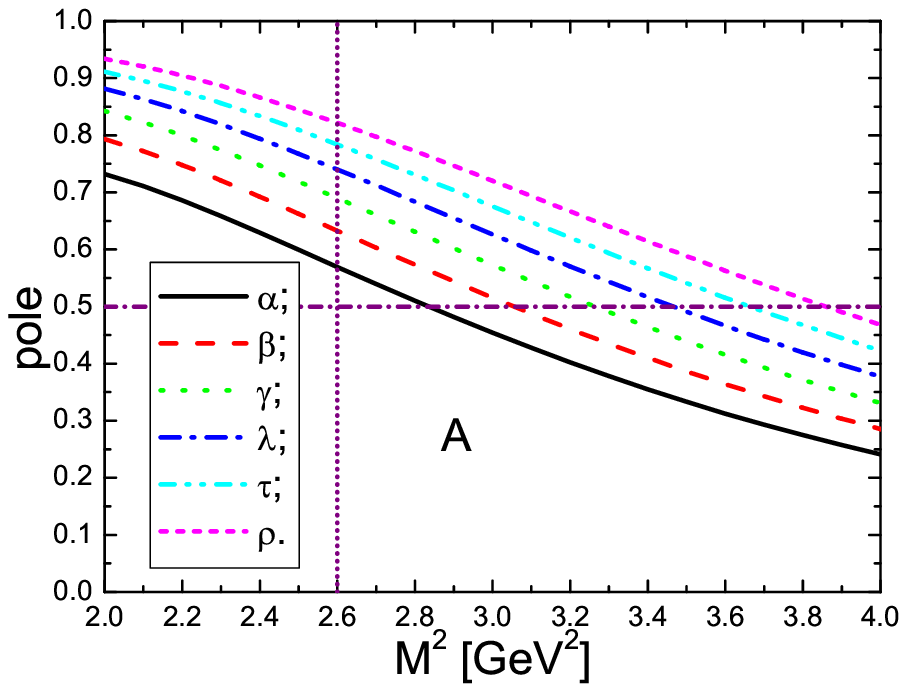}
  \includegraphics[totalheight=5cm,width=6cm]{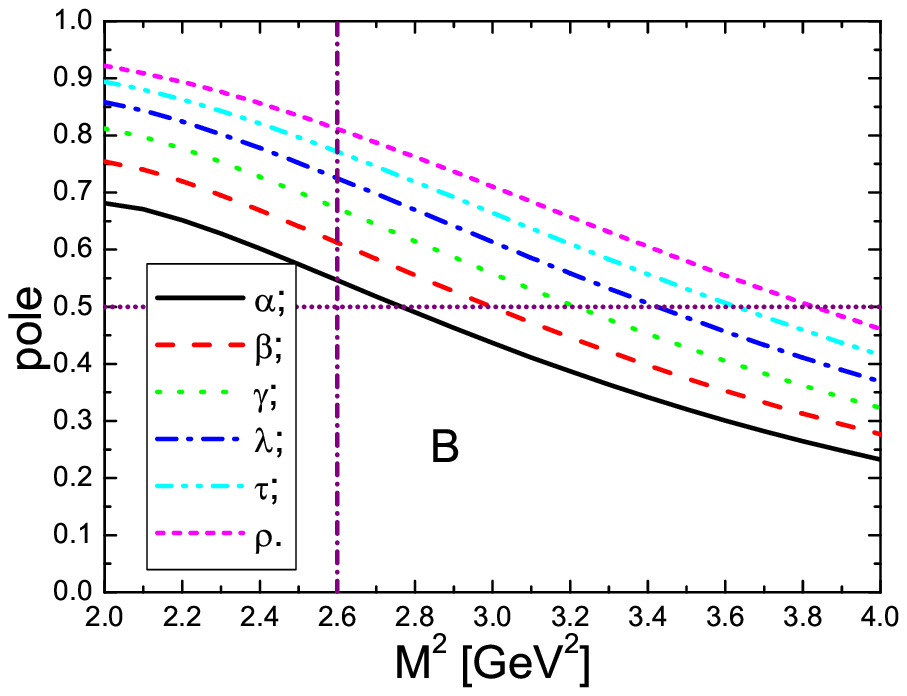}
   \includegraphics[totalheight=5cm,width=6cm]{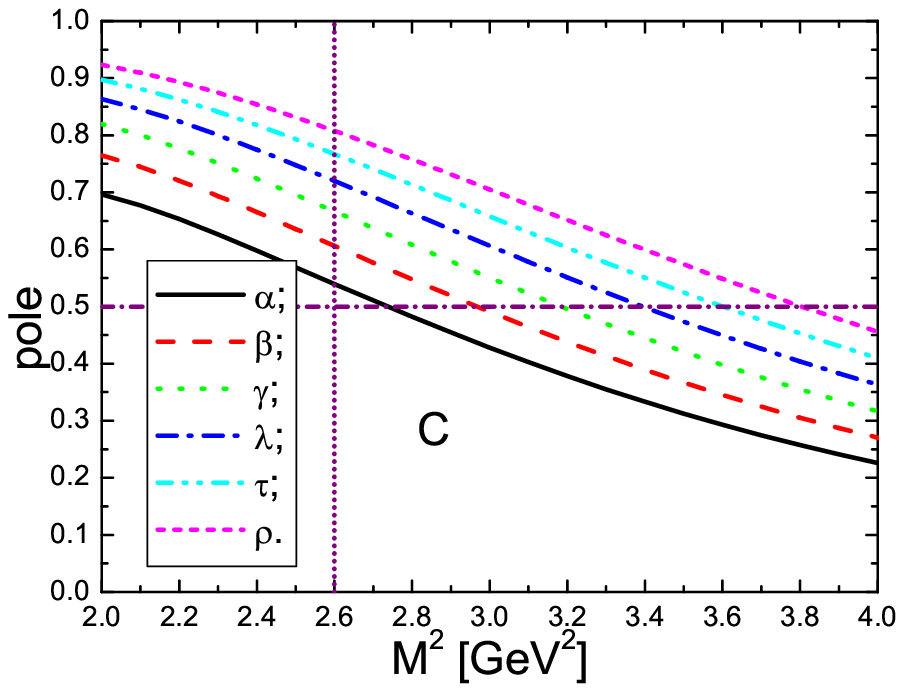}
   \includegraphics[totalheight=5cm,width=6cm]{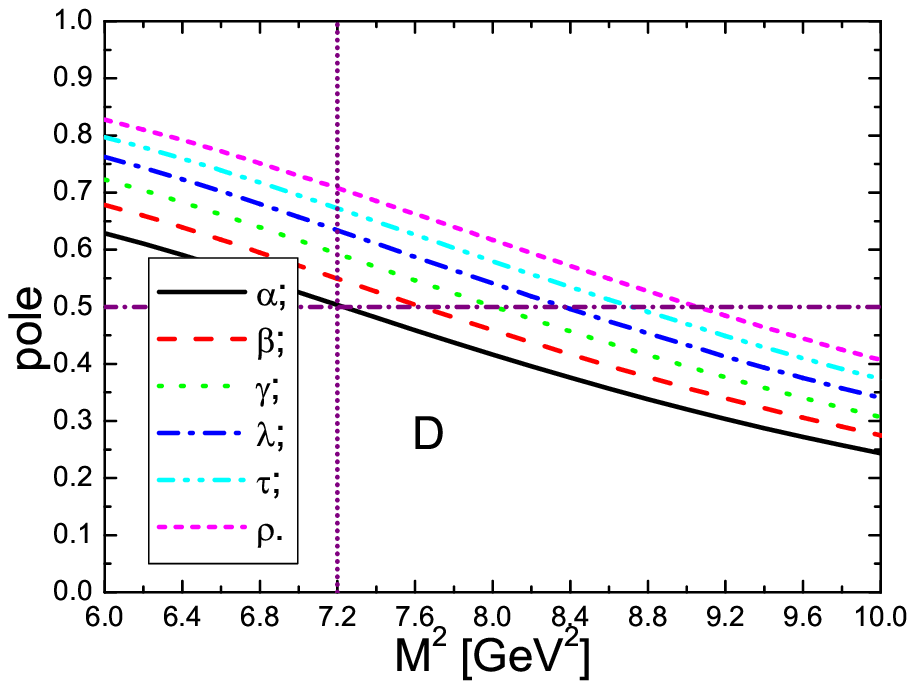}
  \includegraphics[totalheight=5cm,width=6cm]{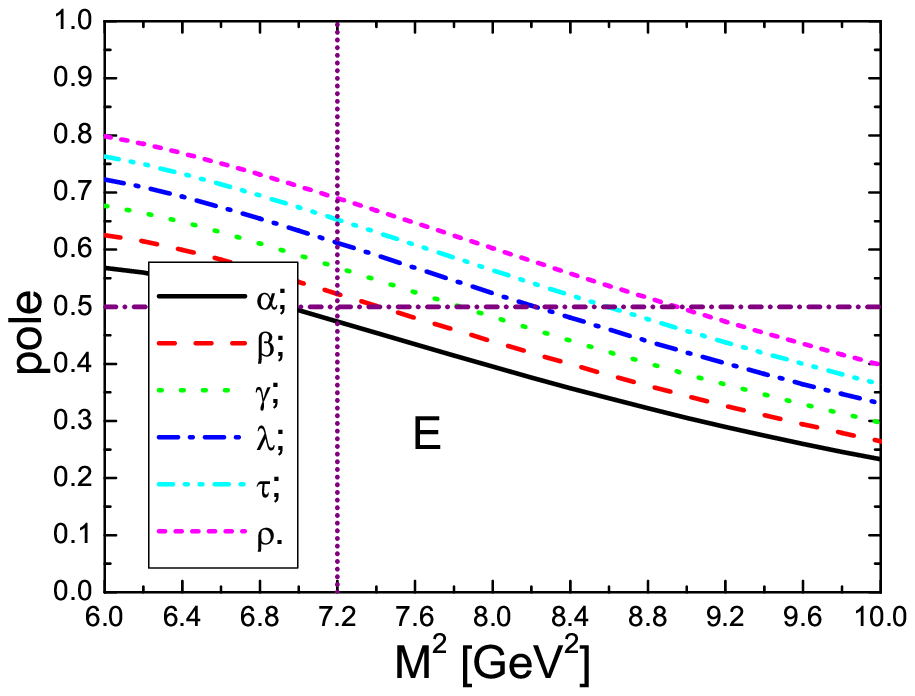}
   \includegraphics[totalheight=5cm,width=6cm]{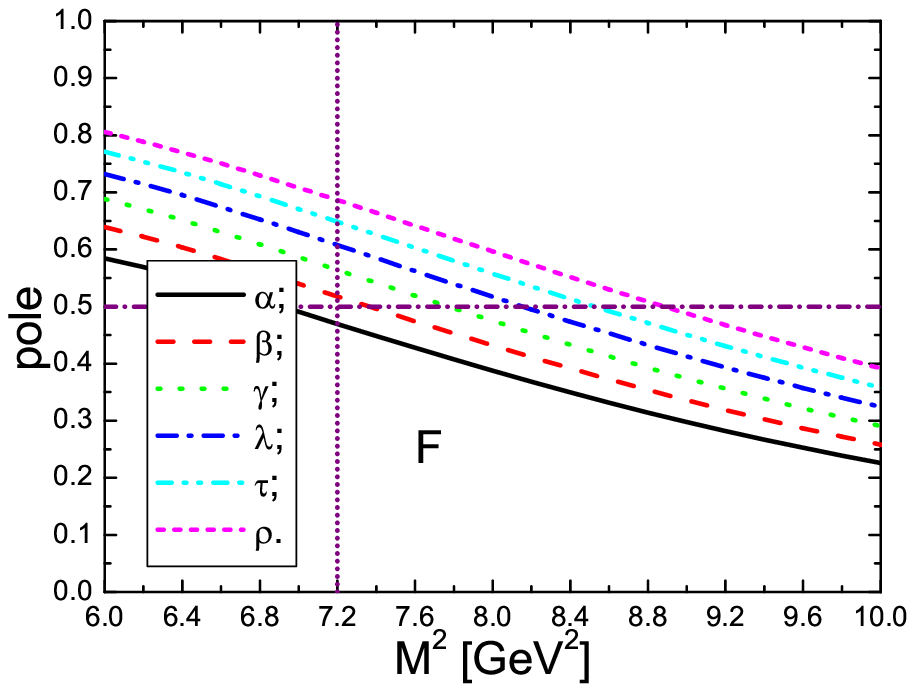}
   \caption{ The contributions of the pole terms with variation of the Borel parameter
   $M^2$. The $A$, $B$, $C$, $D$, $E$ and $F$ denote the channels $c\bar{c}q\bar{q}$,
   $c\bar{c}q\bar{s}$, $c\bar{c}s\bar{s}$, $b\bar{b}q\bar{q}$,
   $b\bar{b}q\bar{s}$ and $b\bar{b}s\bar{s}$ respectively. The notations
   $\alpha$, $\beta$, $\gamma$, $\lambda$, $\tau$ and $\rho$  correspond to the threshold
   parameters $s_0=21\,\rm{GeV}^2$,
   $22\,\rm{GeV}^2$, $23\,\rm{GeV}^2$, $24\,\rm{GeV}^2$, $25\,\rm{GeV}^2$ and $26\,\rm{GeV}^2$
   respectively in the hidden charmed  channels; while they correspond to the threshold
   parameters $s_0=132\,\rm{GeV}^2$,
   $134\,\rm{GeV}^2$, $136\,\rm{GeV}^2$, $138\,\rm{GeV}^2$, $140\,\rm{GeV}^2$ and $142\,\rm{GeV}^2$ respectively
   in the hidden bottom channels.}
\end{figure}

\begin{figure}
 \centering
 \includegraphics[totalheight=5cm,width=6cm]{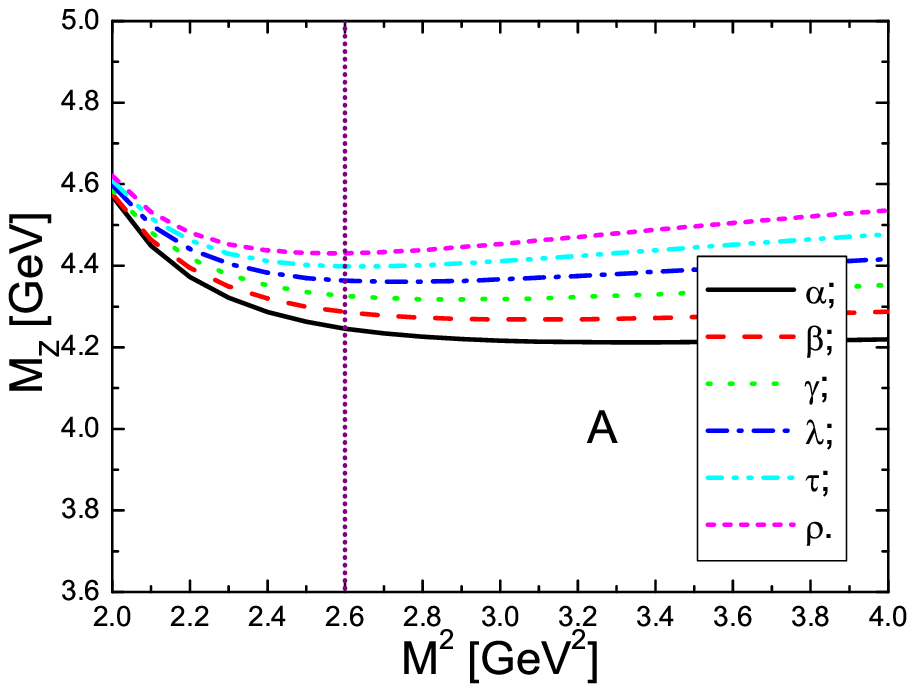}
  \includegraphics[totalheight=5cm,width=6cm]{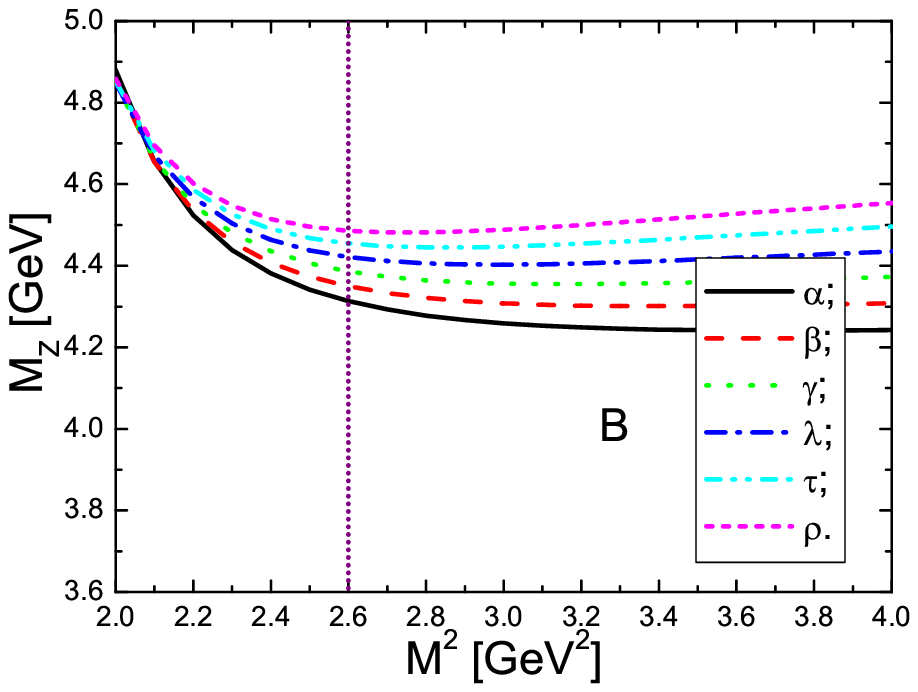}
   \includegraphics[totalheight=5cm,width=6cm]{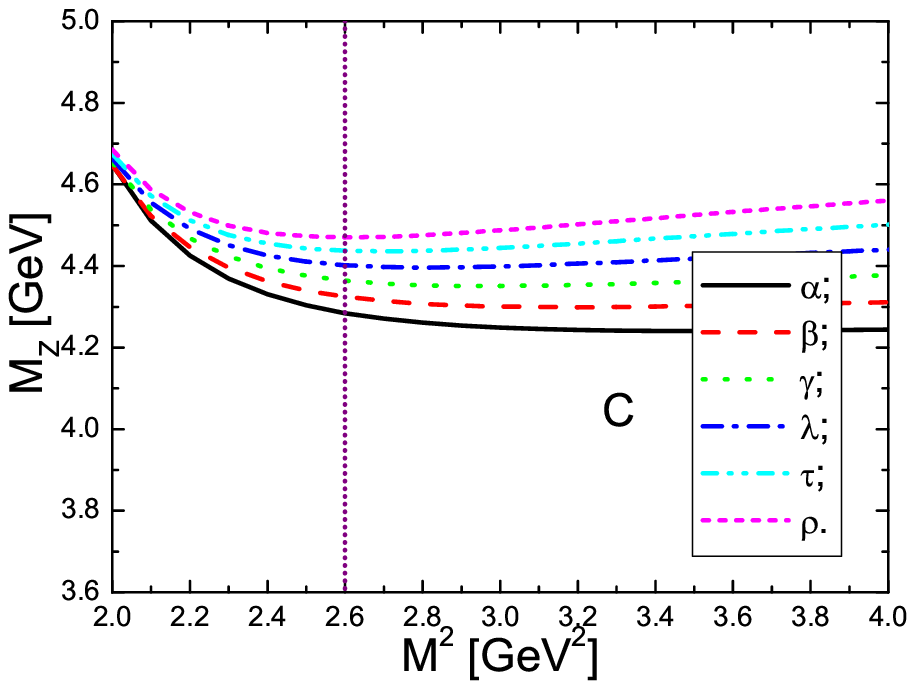}
   \includegraphics[totalheight=5cm,width=6cm]{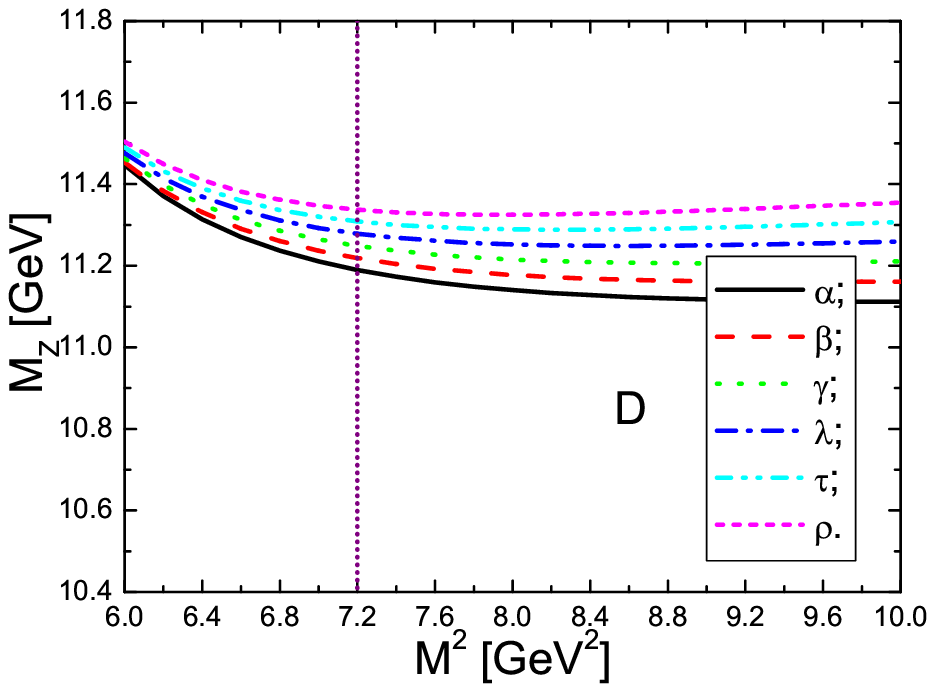}
  \includegraphics[totalheight=5cm,width=6cm]{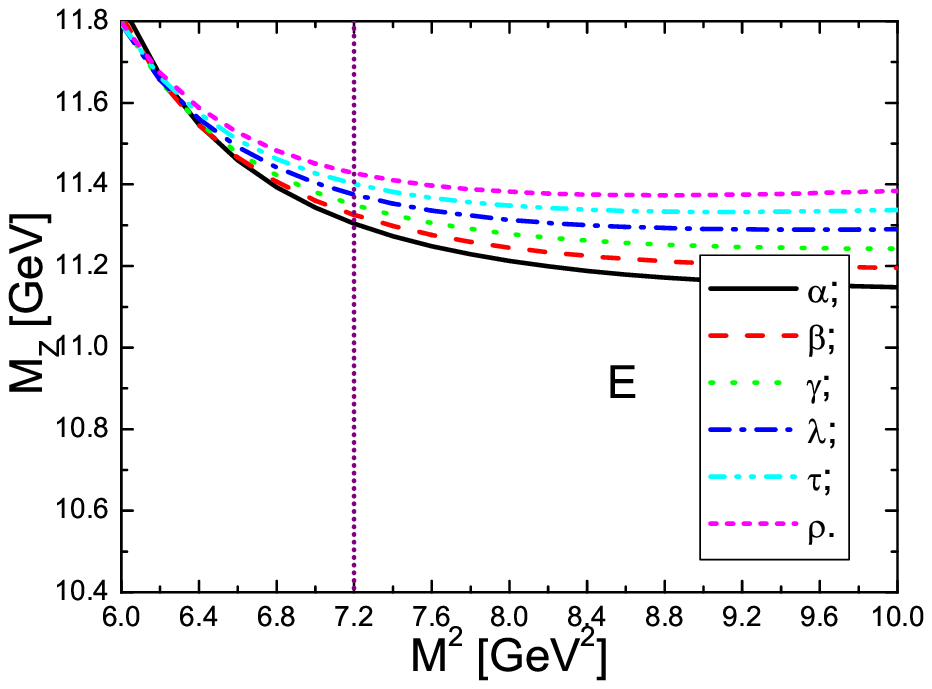}
   \includegraphics[totalheight=5cm,width=6cm]{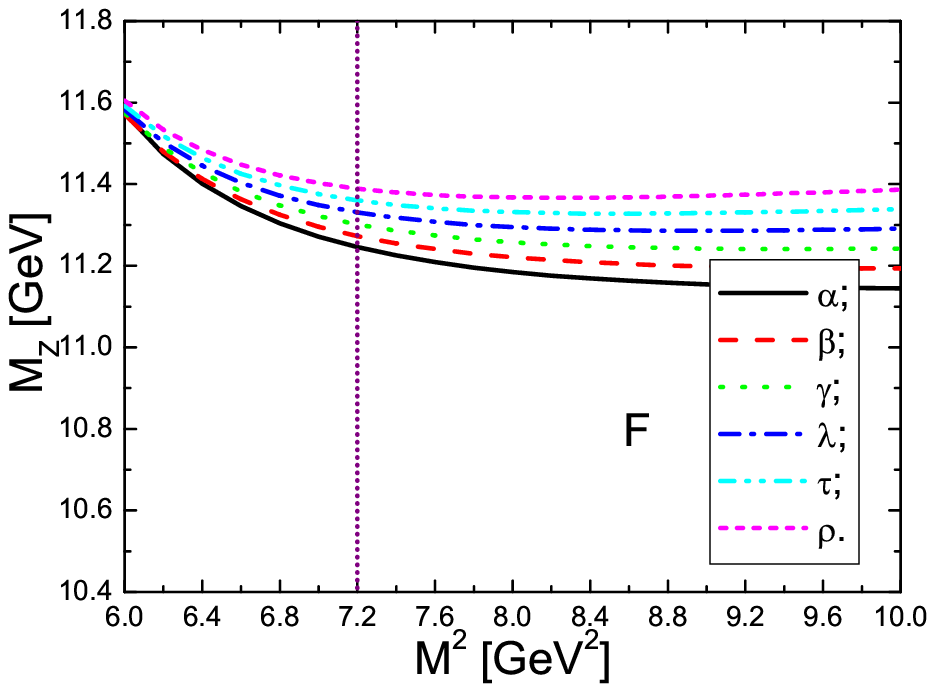}
   \caption{  The masses of the axial-vector tetraquark states with variation of the Borel parameter
   $M^2$. The $A$, $B$, $C$, $D$, $E$ and $F$ denote the channels $c\bar{c}q\bar{q}$,
   $c\bar{c}q\bar{s}$, $c\bar{c}s\bar{s}$, $b\bar{b}q\bar{q}$,
   $b\bar{b}q\bar{s}$ and $b\bar{b}s\bar{s}$ respectively. The notations
   $\alpha$, $\beta$, $\gamma$, $\lambda$, $\tau$ and $\rho$  correspond to the threshold
   parameters $s_0=21\,\rm{GeV}^2$,
   $22\,\rm{GeV}^2$, $23\,\rm{GeV}^2$, $24\,\rm{GeV}^2$, $25\,\rm{GeV}^2$ and $26\,\rm{GeV}^2$
   respectively in the hidden charmed channels; while they correspond to the threshold
   parameters $s_0=132\,\rm{GeV}^2$,
   $134\,\rm{GeV}^2$, $136\,\rm{GeV}^2$, $138\,\rm{GeV}^2$, $140\,\rm{GeV}^2$ and $142\,\rm{GeV}^2$ respectively
   in the hidden bottom channels.}
\end{figure}

\begin{figure}
 \centering
 \includegraphics[totalheight=5cm,width=6cm]{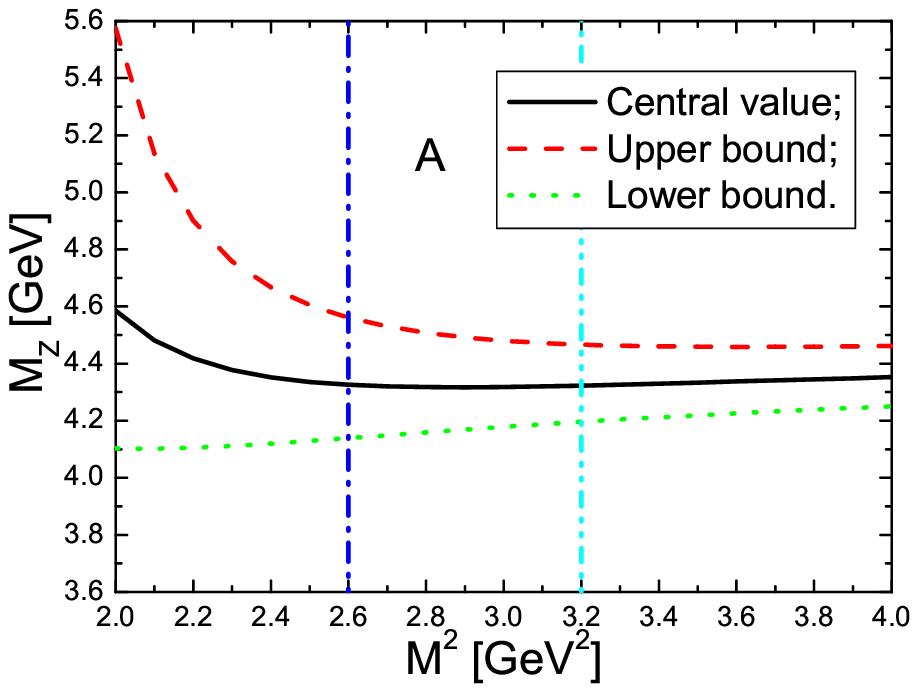}
  \includegraphics[totalheight=5cm,width=6cm]{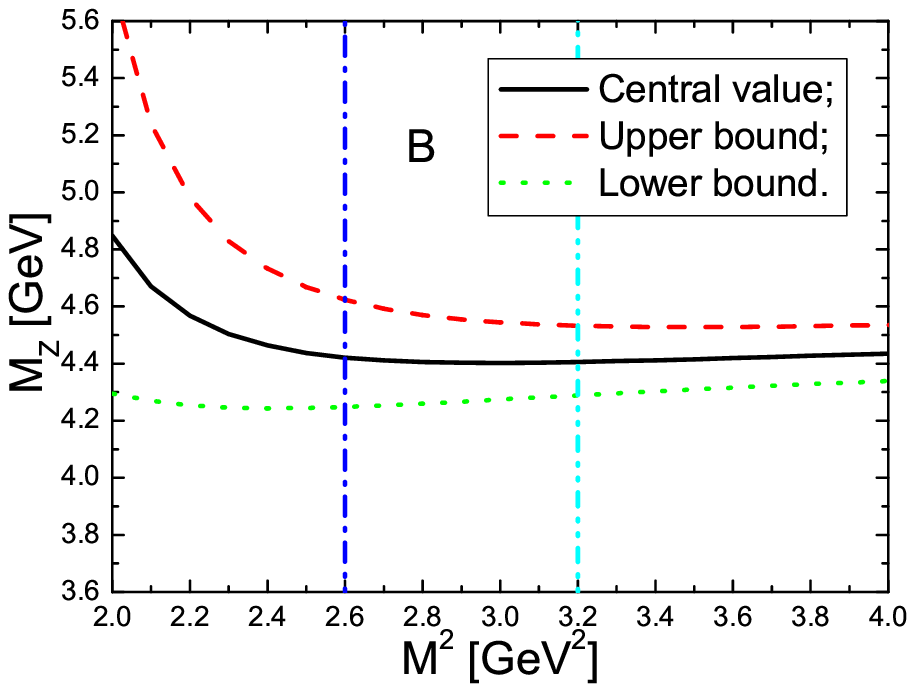}
   \includegraphics[totalheight=5cm,width=6cm]{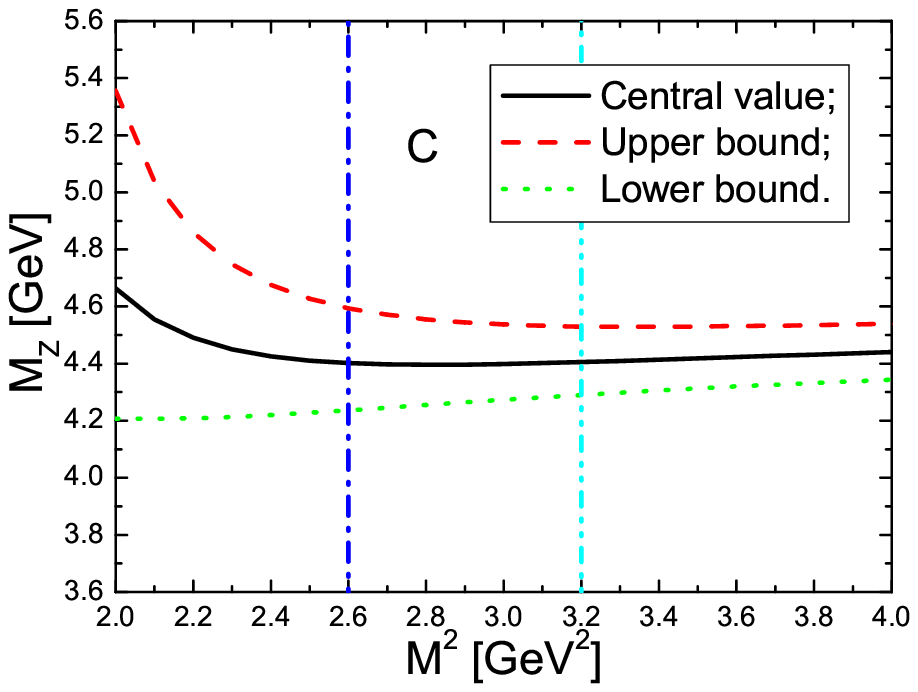}
   \includegraphics[totalheight=5cm,width=6cm]{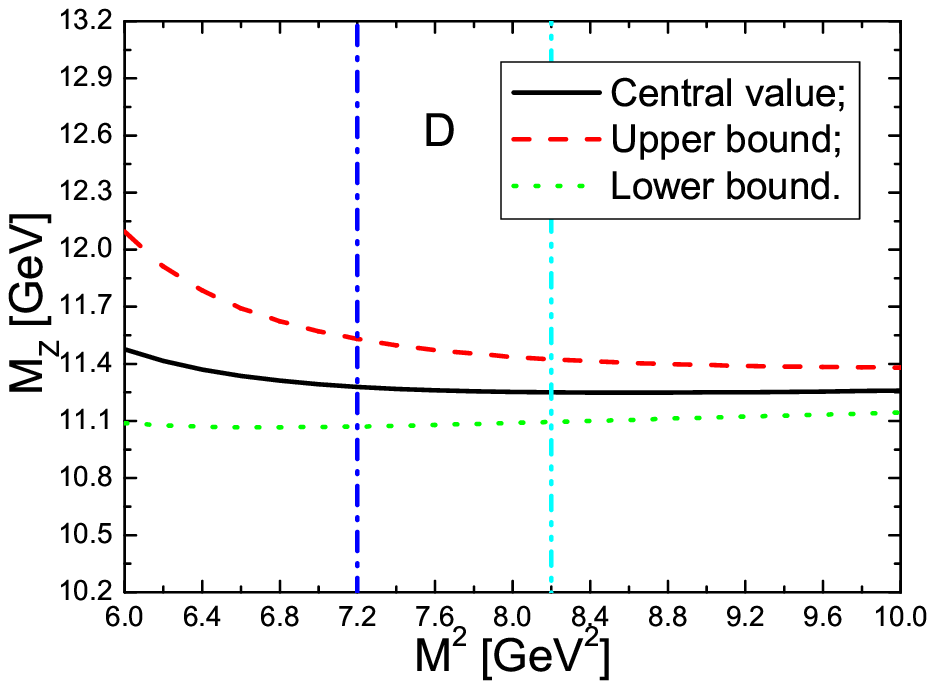}
  \includegraphics[totalheight=5cm,width=6cm]{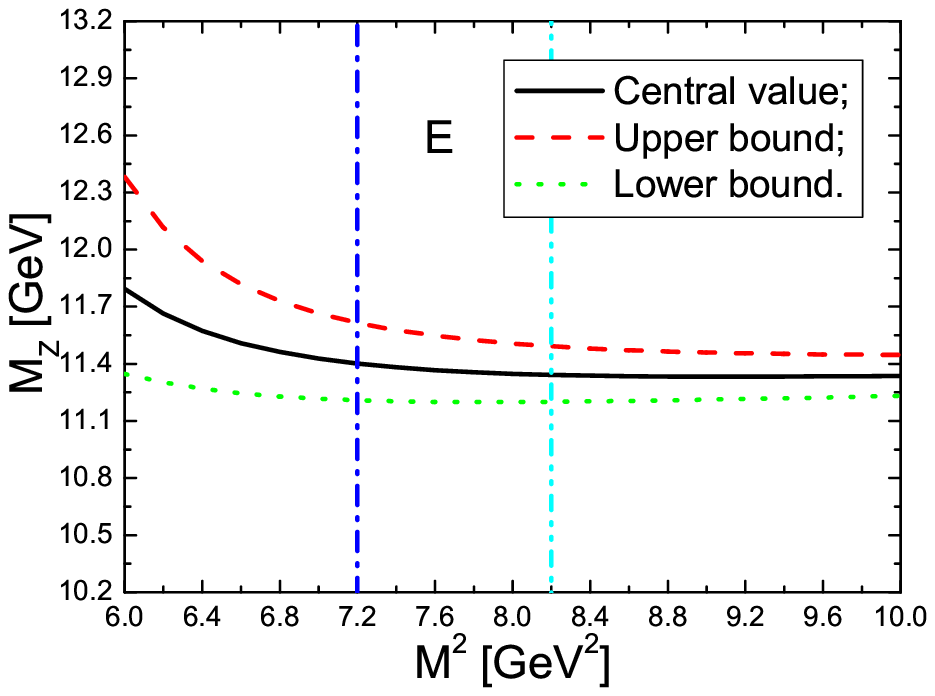}
   \includegraphics[totalheight=5cm,width=6cm]{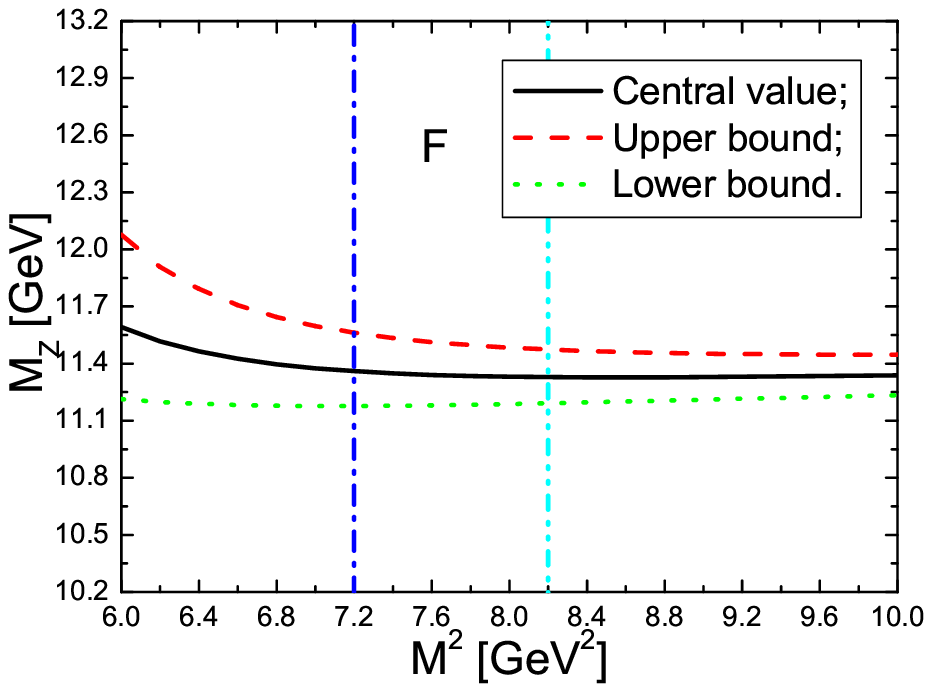}
   \caption{ The masses of the axial-vector tetraquark states with variation of the Borel parameter
   $M^2$. The $A$, $B$, $C$, $D$, $E$ and $F$ denote the channels $c\bar{c}q\bar{q}$,
   $c\bar{c}q\bar{s}$, $c\bar{c}s\bar{s}$, $b\bar{b}q\bar{q}$,
   $b\bar{b}q\bar{s}$ and $b\bar{b}s\bar{s}$ respectively.}
\end{figure}

\begin{figure}
 \centering
 \includegraphics[totalheight=5cm,width=6cm]{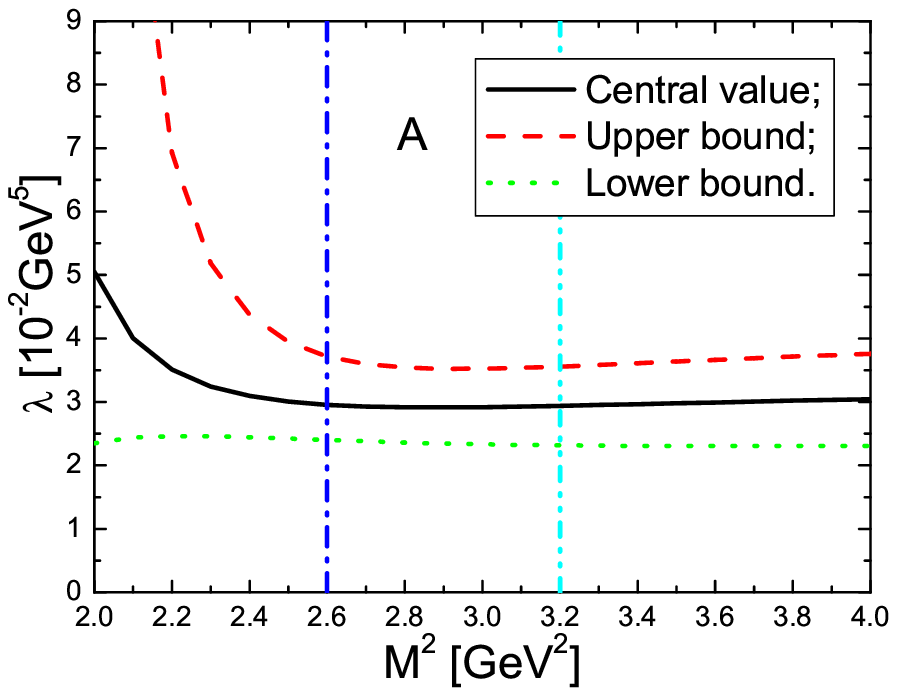}
  \includegraphics[totalheight=5cm,width=6cm]{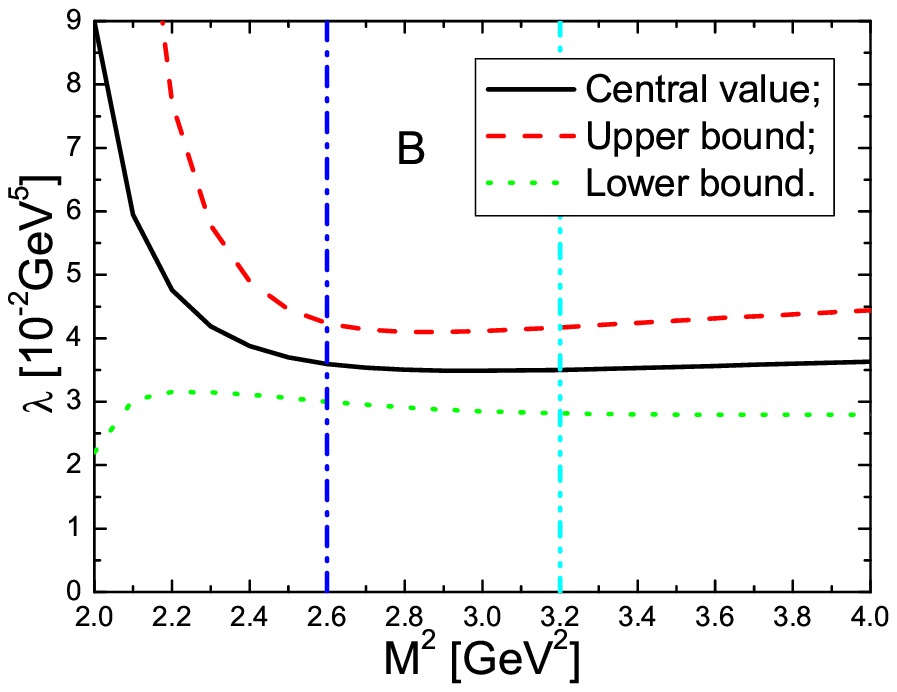}
   \includegraphics[totalheight=5cm,width=6cm]{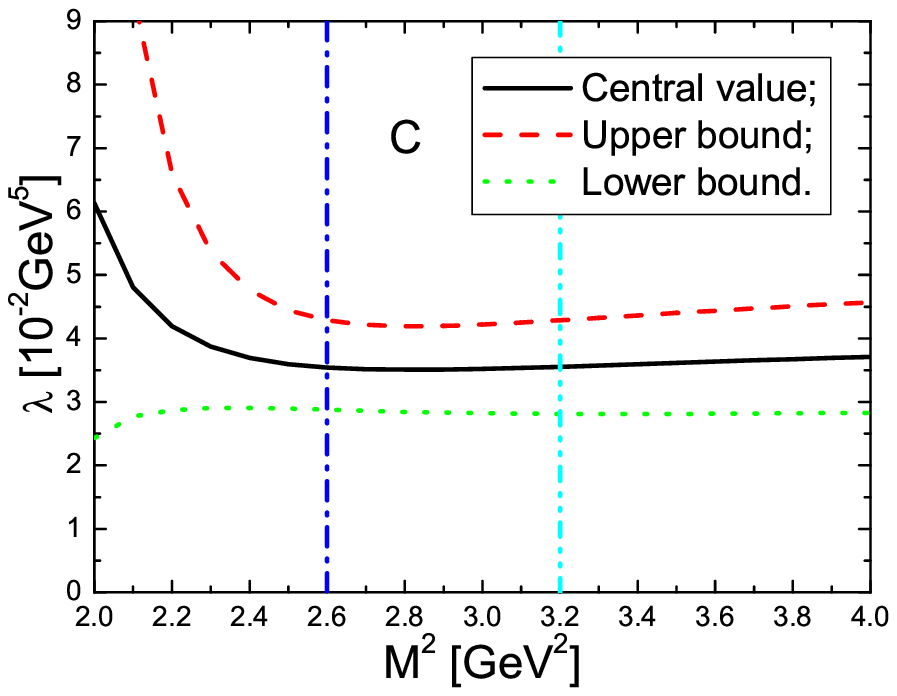}
   \includegraphics[totalheight=5cm,width=6cm]{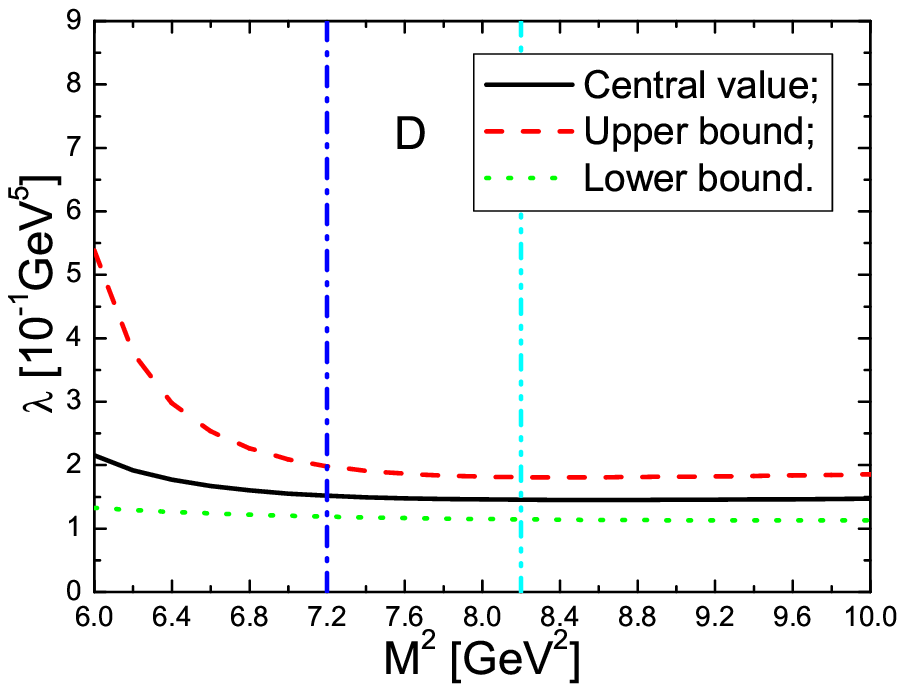}
  \includegraphics[totalheight=5cm,width=6cm]{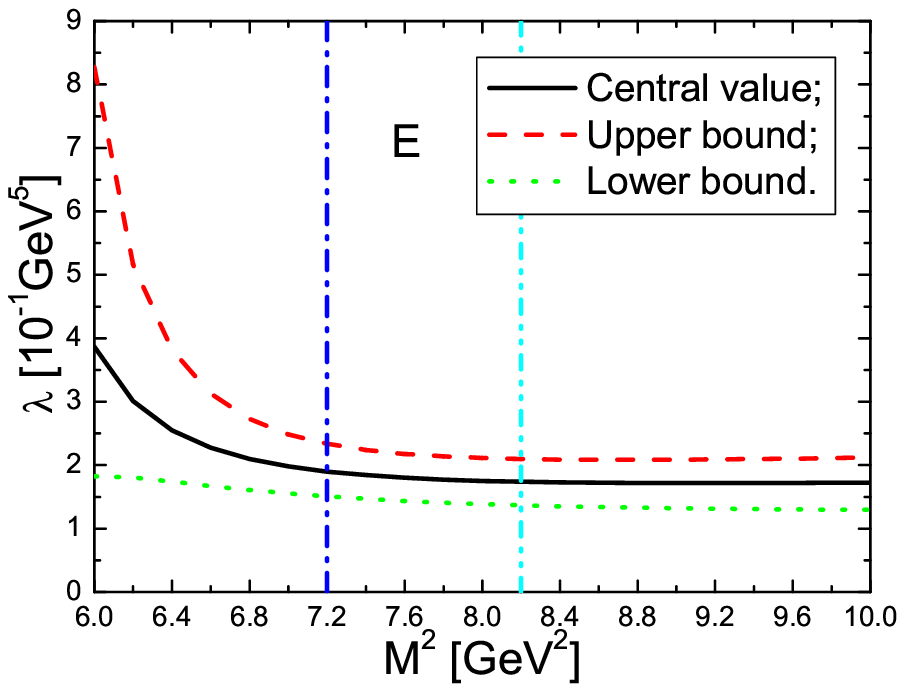}
   \includegraphics[totalheight=5cm,width=6cm]{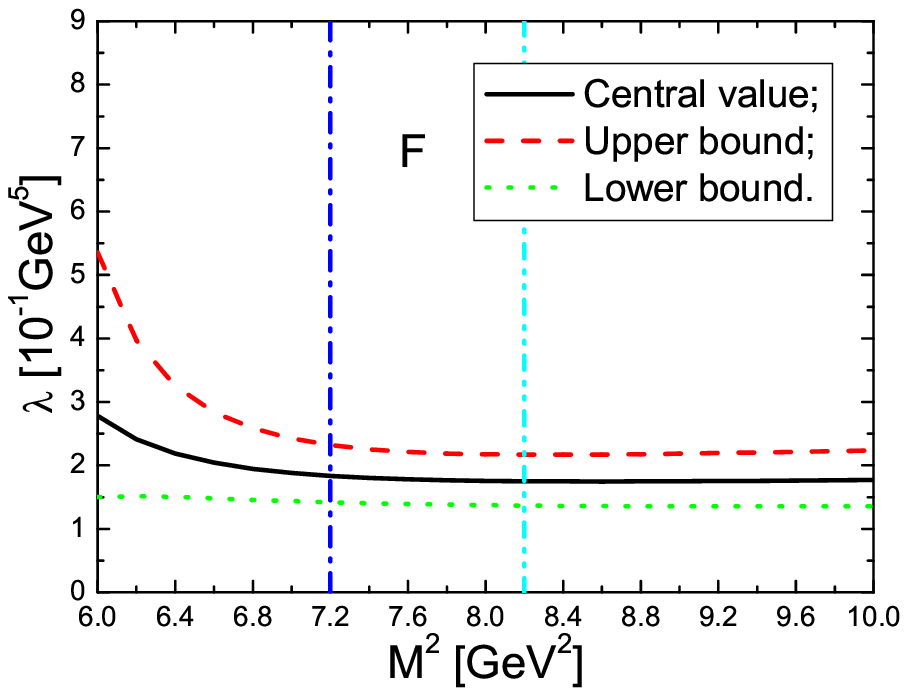}
   \caption{ The pole residues of the axial-vector tetraquark states with variation of the Borel parameter
   $M^2$. The $A$, $B$, $C$, $D$, $E$ and $F$ denote the channels $c\bar{c}q\bar{q}$,
   $c\bar{c}q\bar{s}$, $c\bar{c}s\bar{s}$, $b\bar{b}q\bar{q}$,
   $b\bar{b}q\bar{s}$ and $b\bar{b}s\bar{s}$ respectively.}
\end{figure}

\begin{figure}
 \centering
  \includegraphics[totalheight=7cm,width=8cm]{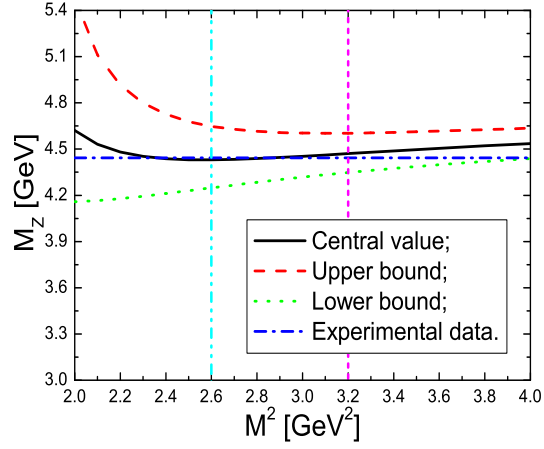}
    \caption{ The mass of  the $Z(4430)$ with variation of the Borel parameter $M^2$.  }
\end{figure}

\begin{figure}
 \centering
  \includegraphics[totalheight=7cm,width=8cm]{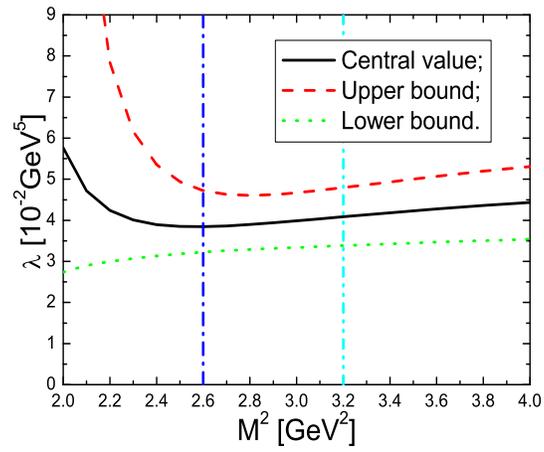}
    \caption{ The pole residue of  the $Z(4430)$ with variation of the Borel parameter $M^2$.  }
\end{figure}

\begin{table}
\begin{center}
\begin{tabular}{|c|c|c|c|c|c|}
\hline\hline tetraquark states & $C\gamma_5-C\gamma_\mu$ &  $\lambda_Z$&$C\gamma_\mu-C\gamma^\mu$&$C\gamma_5-C\gamma_5$\\
\hline
      $c\bar{c}q\bar{q}$  &$4.32\pm0.18$ &$2.92\pm0.60$& $4.36\pm0.18$&$4.37\pm0.18$\\ \hline
       $ c\bar{c}q\bar{s}$& $4.41\pm0.16$&$3.51\pm0.70$ &&$4.39\pm0.16$\\     \hline
      $c\bar{c}s\bar{s} $ &$4.40\pm0.16$&  $3.51\pm0.70$&$4.45\pm0.16$&$4.44\pm0.16$\\      \hline
    $b\bar{b}q\bar{q}$  &$11.27\pm0.20$ & $1.48\pm0.34$&$11.14\pm0.19$&$11.27\pm0.20$\\ \hline
       $ b\bar{b}q\bar{s}$& $11.38\pm0.18$& $1.80\pm0.43$&&$11.33\pm0.16$\\     \hline
      $ b\bar{b}s\bar{s} $ &$11.34\pm0.16$& $1.78\pm0.41$&$11.23\pm0.16$&$11.31\pm0.16$\\      \hline
      $Z(4430)$  &$4.44\pm0.19$ &$3.94\pm0.71$&&\\ \hline
    \hline
\end{tabular}
\end{center}
\caption{ The masses and the pole residues  of the axial-vector
tetraquark states. The masses are in unit of $\rm{GeV}$ and the pole
residues are in unit of $10^{-2}\, \rm{GeV}^5$ and $10^{-1}\,
\rm{GeV}^5$ in the channels $c\bar{c}$ and $b\bar{b}$ respectively.}
\end{table}

The LHCb is a dedicated $b$ and $c$-physics precision experiment at
the LHC (large hadron collider). The LHC will be the world's most
copious  source of the $b$ hadrons, and  a complete spectrum of the
$b$ hadrons will be available through gluon fusion. In proton-proton
collisions at $\sqrt{s}=14\,\rm{TeV}$, the $b\bar{b}$ cross section
is expected to be $\sim 500\mu b$ producing $10^{12}$ $b\bar{b}$
pairs in a standard  year of running at the LHCb operational
luminosity of $2\times10^{32} \rm{cm}^{-2} \rm{sec}^{-1}$
\cite{LHC}. The axial-vector tetraquark states predicted in the
present work may be observed at the LHCb, if they exist indeed. We
can search for the axial-vector hidden charm tetraquark states in
the  $D\bar{D^*}$, $D\bar{D^*_s}$, $D_s\bar{D^*}$, $D_s\bar{D^*_s}$,
$J/\psi \pi$, $J/\psi K$, $J/\psi \eta$, $\psi' \pi$, $\psi' K$,
$\cdots$ invariant mass distributions and search for the
axial-vector hidden bottom tetraquark states in the $B\bar{B^*}$,
$B\bar{B_s^*}$, $B_s\bar{B^*}$, $B_s\bar{B_s^*}$,
 $\Upsilon \pi$, $\Upsilon K$, $\Upsilon
\eta$,  $\Upsilon' \pi$, $\Upsilon' K$, $\Upsilon' \eta$, $\cdots$
invariant mass distributions.

\section{Conclusion}
In this article, we construct the
scalar-diquark-axial-vector-antidiquark type currents to interpolate
the axial-vector tetraquark states, and study the mass spectrum of
the axial-vector hidden charmed and hidden bottom tetraquark states
with the QCD sum rules in a systematic way. In calculations, we take
into account the contributions from the vacuum condensates adding up
to dimension 10 in the operator product expansion, and neglect the
gluon condensates as their  contributions are supposed to be very
small. The mass spectrum   are calculated  by imposing the two
criteria (pole dominance and convergence of the operator product
expansion) of the QCD sum rules. Our numerical result
$M_{Z(4430)}=(4.44\pm0.19)\,\rm{GeV}$ is in excellent agreement with
the experimental data $M_Z=(4433\pm4\pm2)\,\rm{MeV}$ or
$4443^{+15}_{-12}{^{+19}_{-13}}\,\rm{MeV}$ from the Belle
collaboration, which indicates that the $Z^+(4430)$ can be
identified as the axial-vector tetraquark state tentatively.
Considering  the light-flavor $SU(3)$ symmetry and the heavy quark
symmetry, we make predictions for other axial-vector tetraquark
states, the predictions  can be confronted with the experimental
data in the future  at the LHCb or the Fermi-lab Tevatron.

\section*{Appendix}
The spectral densities $\rho_{q\bar{q}}(s)$, $\rho_{q\bar{s}}(s)$
and $\rho_{s\bar{s}}(s)$ at the level of the quark-gluon degrees of
freedom:

\begin{eqnarray}
\rho_{q\bar{q}}(s)&=&\frac{1}{3072 \pi^6}
\int_{\alpha_{i}}^{\alpha_{f}}d\alpha \int_{\beta_{i}}^{1-\alpha}
d\beta
\alpha\beta(1-\alpha-\beta)^3(s-\widetilde{m}^2_Q)^2(35s^2-26s\widetilde{m}^2_Q+3\widetilde{m}^4_Q)
\nonumber \\
&&-\frac{ m_Q\langle \bar{q}q\rangle}{32 \pi^4}
\int_{\alpha_{i}}^{\alpha_{f}}d\alpha \int_{\beta_{i}}^{1-\alpha}
d\beta
(1-\alpha-\beta)(s-\widetilde{m}^2_Q) \left[(3\alpha+4\beta) s-(\alpha+2\beta)\widetilde{m}^2_Q\right] \nonumber\\
&& +\frac{ m_Q\langle \bar{q}g_s\sigma Gq\rangle}{64 \pi^4}
\int_{\alpha_{i}}^{\alpha_{f}}d\alpha \int_{\beta_{i}}^{1-\alpha}
d\beta
\left[(2\alpha+3\beta)s-(\alpha+2\beta)\widetilde{m}^2_Q) \right] \nonumber\\
&&+\frac{m_Q^2\langle \bar{q}q\rangle^2}{12 \pi^2}
\int_{\alpha_{i}}^{\alpha_{f}} d\alpha
-\frac{m_Q^2\langle\bar{q}q\rangle\langle\bar{q}g_s \sigma
Gq\rangle}{24\pi^2}\int_{\alpha_{i}}^{\alpha_{f}} d\alpha \left[
1+\frac{s}{M^2}\right]\delta(s-\widetilde{\widetilde{m}}_Q^2)
\nonumber \\
&&+\frac{m_Q^2\langle\bar{q}g_s \sigma
Gq\rangle^2}{192\pi^2M^6}\int_{\alpha_{i}}^{\alpha_{f}} d\alpha
s^2\delta(s-\widetilde{\widetilde{m}}_Q^2)\, ,
\end{eqnarray}

\begin{eqnarray}
\rho_{q\bar{s}}(s)&=&\frac{1}{3072 \pi^6}
\int_{\alpha_{i}}^{\alpha_{f}}d\alpha \int_{\beta_{i}}^{1-\alpha}
d\beta
\alpha\beta(1-\alpha-\beta)^3(s-\widetilde{m}^2_Q)^2(35s^2-26s\widetilde{m}^2_Q+3\widetilde{m}^4_Q)
\nonumber \\
&&+\frac{ m_sm_Q}{256 \pi^6} \int_{\alpha_{i}}^{\alpha_{f}}d\alpha
\int_{\beta_{i}}^{1-\alpha} d\beta \beta
(1-\alpha-\beta)^2(s-\widetilde{m}^2_Q)^2(5s-2\widetilde{m}^2_Q)   \nonumber\\
&&+\frac{ m_s\langle \bar{s}s\rangle}{64 \pi^4}
\int_{\alpha_{i}}^{\alpha_{f}}d\alpha \int_{\beta_{i}}^{1-\alpha}
d\beta \alpha \beta
(1-\alpha-\beta)(15s^2-16s\widetilde{m}^2_Q+3\widetilde{m}^4_Q)   \nonumber\\
&&+\frac{ m_Q\langle \bar{q}q\rangle}{32 \pi^4}
\int_{\alpha_{i}}^{\alpha_{f}}d\alpha \int_{\beta_{i}}^{1-\alpha}
d\beta \alpha
(1-\alpha-\beta)(s-\widetilde{m}^2_Q) (\widetilde{m}^2_Q-3s) \nonumber\\
&&+\frac{ m_Q\langle \bar{s}s\rangle}{16 \pi^4}
\int_{\alpha_{i}}^{\alpha_{f}}d\alpha \int_{\beta_{i}}^{1-\alpha}
d\beta \beta
(1-\alpha-\beta) (s-\widetilde{m}^2_Q) (\widetilde{m}^2_Q-2s) \nonumber\\
&& +\frac{ m_Q\langle \bar{q}g_s\sigma Gq\rangle}{64 \pi^4}
\int_{\alpha_{i}}^{\alpha_{f}}d\alpha \int_{\beta_{i}}^{1-\alpha}
d\beta \alpha
(2s-\widetilde{m}^2_Q)  \nonumber\\
&& + \frac{ m_Q\langle \bar{s}g_s\sigma Gs\rangle}{64 \pi^4}
\int_{\alpha_{i}}^{\alpha_{f}}d\alpha \int_{\beta_{i}}^{1-\alpha}
d\beta \beta
(3s-2\widetilde{m}^2_Q)  \nonumber\\
&&-\frac{ m_s\langle \bar{s}g_s\sigma Gs\rangle}{192 \pi^4}
\int_{\alpha_{i}}^{\alpha_{f}}d\alpha \int_{\beta_{i}}^{1-\alpha}
d\beta \alpha \beta
\left[8s-3\widetilde{m}^2_Q+s^2\delta(s-\widetilde{m}^2_Q)\right]   \nonumber\\
&& -\frac{ m_sm_Q^2\langle \bar{q}q\rangle}{16 \pi^4}
\int_{\alpha_{i}}^{\alpha_{f}}d\alpha \int_{\beta_{i}}^{1-\alpha}
d\beta
(s-\widetilde{m}^2_Q)  \nonumber\\
&&+\frac{m_Q^2\langle \bar{q}q\rangle \langle \bar{s}s\rangle}{12
\pi^2} \int_{\alpha_{i}}^{\alpha_{f}} d\alpha +\frac{m_sm_Q^2\langle
\bar{q}g_s\sigma Gq\rangle }{64 \pi^4}
\int_{\alpha_{i}}^{\alpha_{f}} d\alpha \nonumber\\
&&-\frac{m_sm_Q\langle \bar{q}q\rangle \langle \bar{s}s\rangle}{24
\pi^2} \int_{\alpha_{i}}^{\alpha_{f}} d\alpha \alpha
\left[1+s\delta(s-\widetilde{\widetilde{m}}^2_Q) \right]\nonumber\\
&&-\frac{m_Q^2\left[\langle\bar{q}q\rangle\langle\bar{s}g_s \sigma
Gs\rangle+\langle\bar{s}s\rangle\langle\bar{q}g_s \sigma
Gq\rangle\right]}{48\pi^2}\int_{\alpha_{i}}^{\alpha_{f}} d\alpha
\left[ 1+\frac{s}{M^2}\right]\delta(s-\widetilde{\widetilde{m}}_Q^2)
\nonumber \\
&&+\frac{m_sm_Q\left[2\langle\bar{q}q\rangle\langle\bar{s}g_s \sigma
Gs\rangle+3\langle\bar{s}s\rangle\langle\bar{q}g_s \sigma
Gq\rangle\right]}{288\pi^2M^2}\int_{\alpha_{i}}^{\alpha_{f}} d\alpha
\alpha\left[
s-\frac{s^2}{M^2}\right]\delta(s-\widetilde{\widetilde{m}}_Q^2)
\nonumber \\
&&+\frac{m_Q^2\langle\bar{q}g_s \sigma Gq\rangle\langle\bar{s}g_s
\sigma Gs\rangle}{192\pi^2M^6}\int_{\alpha_{i}}^{\alpha_{f}} d\alpha
s^2\delta(s-\widetilde{\widetilde{m}}_Q^2) \, ,
\end{eqnarray}

\begin{eqnarray}
\rho_{s\bar{s}}(s)&=&\frac{1}{3072 \pi^6}
\int_{\alpha_{i}}^{\alpha_{f}}d\alpha \int_{\beta_{i}}^{1-\alpha}
d\beta
\alpha\beta(1-\alpha-\beta)^3(s-\widetilde{m}^2_Q)^2(35s^2-26s\widetilde{m}^2_Q+3\widetilde{m}^4_Q)
\nonumber \\
&&+ \frac{ m_sm_Q}{256 \pi^6} \int_{\alpha_{i}}^{\alpha_{f}}d\alpha
\int_{\beta_{i}}^{1-\alpha} d\beta
(1-\alpha-\beta)^2(s-\widetilde{m}^2_Q)^2\left[(4\alpha+5\beta)s-(\alpha+2\beta)\widetilde{m}^2_Q \right]   \nonumber\\
&&+\frac{ m_s\langle \bar{s}s\rangle}{32 \pi^4}
\int_{\alpha_{i}}^{\alpha_{f}}d\alpha \int_{\beta_{i}}^{1-\alpha}
d\beta \alpha \beta
(1-\alpha-\beta)(15s^2-16s\widetilde{m}^2_Q+3\widetilde{m}^4_Q)   \nonumber\\
&&-\frac{ m_Q\langle \bar{s}s\rangle}{32 \pi^4}
\int_{\alpha_{i}}^{\alpha_{f}}d\alpha \int_{\beta_{i}}^{1-\alpha}
d\beta
(1-\alpha-\beta) (s-\widetilde{m}^2_Q) \left[(3\alpha+4\beta)s-(\alpha+2\beta)\widetilde{m}^2_Q \right]\nonumber\\
&& +\frac{ m_Q\langle \bar{s}g_s\sigma Gs\rangle}{64 \pi^4}
\int_{\alpha_{i}}^{\alpha_{f}}d\alpha \int_{\beta_{i}}^{1-\alpha}
d\beta
\left[(2\alpha+3\beta)s-(\alpha+2\beta)\widetilde{m}^2_Q \right]  \nonumber\\
&&-\frac{ m_s\langle \bar{s}g_s\sigma Gs\rangle}{96 \pi^4}
\int_{\alpha_{i}}^{\alpha_{f}}d\alpha \int_{\beta_{i}}^{1-\alpha}
d\beta \alpha \beta
\left[8s-3\widetilde{m}^2_Q+s^2\delta(s-\widetilde{m}^2_Q)\right]   \nonumber\\
&& -\frac{ m_sm_Q^2\langle \bar{s}s\rangle}{8 \pi^4}
\int_{\alpha_{i}}^{\alpha_{f}}d\alpha \int_{\beta_{i}}^{1-\alpha}
d\beta
(s-\widetilde{m}^2_Q)  \nonumber\\
&&+\frac{m_Q^2  \langle \bar{s}s\rangle^2}{12 \pi^2}
\int_{\alpha_{i}}^{\alpha_{f}} d\alpha +\frac{m_sm_Q^2\langle
\bar{s}g_s\sigma Gs\rangle }{32 \pi^4}
\int_{\alpha_{i}}^{\alpha_{f}} d\alpha \nonumber\\
&&-\frac{m_sm_Q \langle \bar{s}s\rangle^2}{24 \pi^2}
\int_{\alpha_{i}}^{\alpha_{f}} d\alpha (1-\alpha)
\left[2+s\delta(s-\widetilde{\widetilde{m}}^2_Q) \right]\nonumber\\
&&-\frac{m_sm_Q \langle \bar{s}s\rangle^2}{24 \pi^2}
\int_{\alpha_{i}}^{\alpha_{f}} d\alpha \alpha
\left[1+s\delta(s-\widetilde{\widetilde{m}}^2_Q) \right]\nonumber\\
&&-\frac{m_Q^2\langle\bar{s}s\rangle\langle\bar{s}g_s \sigma
Gs\rangle}{24\pi^2}\int_{\alpha_{i}}^{\alpha_{f}} d\alpha \left[
1+\frac{s}{M^2}\right]\delta(s-\widetilde{\widetilde{m}}_Q^2)\nonumber\\
&&+\frac{5m_sm_Q\langle\bar{s}s\rangle\langle\bar{s}g_s \sigma
Gs\rangle}{288\pi^2M^2}\int_{\alpha_{i}}^{\alpha_{f}} d\alpha \alpha
\left[
s-\frac{s^2}{M^2}\right]\delta(s-\widetilde{\widetilde{m}}_Q^2)
\nonumber\\
&&+\frac{5m_sm_Q\langle\bar{s}s\rangle\langle\bar{s}g_s \sigma
Gs\rangle}{144\pi^2}\int_{\alpha_{i}}^{\alpha_{f}} d\alpha
(1-\alpha) \left[1+
\frac{s}{M^2}+\frac{s^2}{2M^4}\right]\delta(s-\widetilde{\widetilde{m}}_Q^2)
\nonumber \\
&&+\frac{m_Q^2\langle\bar{s}g_s \sigma
Gs\rangle^2}{192\pi^2M^6}\int_{\alpha_{i}}^{\alpha_{f}} d\alpha
s^2\delta(s-\widetilde{\widetilde{m}}_Q^2)\, ,
\end{eqnarray}

\section*{Acknowledgements}
This  work is supported by National Natural Science Foundation of
China, Grant Number 10775051, and Program for New Century Excellent
Talents in University, Grant Number NCET-07-0282, and the
Fundamental Research Funds for the Central Universities.


\begin{thebibliography}{99}


\bibitem{review1} E. S. Swanson, Phys. Rept. {\bf 429} (2006) 243.

\bibitem{review2} E. Klempt and A. Zaitsev, Phys. Rept. {\bf 454} (2007) 1.

\bibitem{review3} M. B. Voloshin, Prog. Part. Nucl. Phys. {\bf 61} (2008) 455.

\bibitem{review4} S. Godfrey and  S. L. Olsen, Ann. Rev. Nucl. Part. Sci. {\bf 58} (2008) 51.

\bibitem{Recent-review} N. Drenska, R. Faccini, F. Piccinini, A. Polosa, F. Renga and C. Sabelli, arXiv:1006.2741.


\bibitem{Olsen2009}  S. L. Olsen,   Nucl. Phys. {\bf A827} (2009) 53C.

\bibitem{Belle-z4430} S. K. Choi et al, Phys. Rev. Lett. {\bf 100} (2008) 142001.

\bibitem{Belle-z4430-PRD} R. Mizuk et al,  Phys. Rev. {\bf D80} (2009) 031104.


\bibitem{Babar0811} B. Aubert et al,  Phys. Rev. {\bf D79} (2009) 112001.

\bibitem{Belle-chipi}  R. Mizuk  et al,  Phys. Rev. {\bf D78} (2008) 072004.

\bibitem{Voloshin0803} S. Dubynskiy  and M. B. Voloshin,  Phys. Lett. {\bf B666} (2008) 344.

\bibitem{Rosner07} J. L. Rosner, Phys. Rev. {\bf D76} (2007) 114002.


\bibitem{Meng07} C. Meng and K. T. Chao, arXiv:0708.4222.

\bibitem{Lee07-sum} S. H. Lee, A. Mihara, F. S. Navarra and M. Nielsen, Phys. Lett. {\bf B661} (2008) 28.

\bibitem{Liu07} X. Liu, Y. R. Liu, W. Z. Deng and S. L. Zhu, Phys. Rev. {\bf D77} (2008) 034003.


\bibitem{Ding07} G. J. Ding, arXiv:0711.1485.

\bibitem{Braaten07} E. Braaten and M. Lu, Phys. Rev. {\bf D79} (2009) 051503.


\bibitem{Liu0803} X. Liu, Y. R. Liu, W. Z. Deng and S. L. Zhu, Phys. Rev. {\bf D77} (2008) 094015.
\bibitem{Meng0905} G. Z. Meng et al,   Phys. Rev. {\bf D80} (2009) 034503.
\bibitem{Ding0805} G. J. Ding, W. Huang, J. F. Liu and M. L. Yan,  Phys. Rev. {\bf D79} (2009) 034026.



\bibitem{Maiani07} L. Maiani, A. D. Polosa and V. Riquer, arXiv:0708.3997.

\bibitem{Gershtein07} S. S. Gershtein, A. K. Likhoded and G. P. Pronko, arXiv:0709.2058.

\bibitem{Li07} Y. Li, C. D. Lu and W. Wang, Phys. Rev. {\bf D77} (2008) 054001.

\bibitem{Liu08} X. H. Liu,  Q. Zhao and F. E. Close, Phys. Rev. {\bf D77} (2008) 094005.
\bibitem{Cheung0709} K. M. Cheung, W. Y. Keung and T. C. Yuan, Phys. Rev. {\bf D76} (2007) 117501.

\bibitem{Bracco0807} M. E. Bracco,  S. H. Lee,  M. Nielsen and R. Rodrigues da Silva,  Phys. Lett. {\bf B671} (2009) 240.



\bibitem{Bugg07} D. V. Bugg, arXiv:0709.1254.
\bibitem{Matsuki0805} T. Matsuki, T. Morii and K. Sudoh, Phys. Lett. {\bf B669} (2008) 156.
\bibitem{Danilkin0902} I. V. Danilkin and P. Y. Kulikov, JETP Lett. {\bf 89} (2009) 390.

\bibitem{Wang0807} Z. G. Wang, Eur. Phys. J. {\bf C59} (2009) 675.

\bibitem{Wang08072} Z. G. Wang, Eur. Phys. J. {\bf C62} (2009) 375.

\bibitem{WangScalar} Z. G. Wang, Phys. Rev. {\bf D79} (2009) 094027.

\bibitem{WangScalar-2} Z. G. Wang,  Eur. Phys. J. {\bf C67} (2010) 411.


\bibitem{WangVector} Z. G. Wang, J. Phys. {\bf G36} (2009) 085002.

\bibitem{Zhu2010} W. Chen and S. L. Zhu,  Phys. Rev. {\bf D81} (2010) 105018.


\bibitem{LHC}  G. Kane and A. Pierce, "Perspectives On LHC Physics",
World Scientific Publishing Company,  2008.

\bibitem{Jaffe2003} R. L. Jaffe and  F. Wilczek, Phys. Rev. Lett. {\bf 91} (2003) 232003.


\bibitem{Jaffe2004} R. L. Jaffe, Phys. Rept. {\bf 409} (2005) 1.

\bibitem{GI1} A. De Rujula, H. Georgi and S. L. Glashow, Phys. Rev.  {\bf D12}
(1975) 147.

\bibitem{GI2} T. DeGrand, R. L. Jaffe, K. Johnson and J. E. Kiskis,
Phys.  Rev.  {\bf D12} (1975) 2060.

\bibitem{SVZ79}  M. A. Shifman, A. I. Vainshtein and V. I. Zakharov, Nucl. Phys. {\bf B147} (1979) 385.

\bibitem{Reinders85} L. J. Reinders, H. Rubinstein and S. Yazaki, Phys. Rept. {\bf 127} (1985) 1.

\bibitem{Wang1} Z. G. Wang, Nucl. Phys. {\bf A791} (2007) 106.


\bibitem{Wang2} Z. G. Wang, W. M. Yang and S. L. Wan, J. Phys. {\bf G31} (2005) 971.

\bibitem{Wang0904} Z. G. Wang, Eur. Phys. J. {\bf C63} (2009) 115.

\bibitem{Wang0907} Z. G. Wang, Z. C. Liu and X. H. Zhang,  Eur. Phys. J. {\bf C64} (2009) 373.

\bibitem{WangZhang1} Z. G. Wang and X. H. Zhang, Commun. Theor. Phys. {\bf 54} (2010) 323.

\bibitem{WangZhang2} Z. G. Wang and X. H. Zhang,  Eur. Phys. J. {\bf C66} (2010) 419.

\bibitem{Cotugno-4} G. Cotugno,  R. Faccini, A. D. Polosa and C. Sabelli, Phys. Rev. Lett. {\bf 104} (2010) 132005.



\bibitem{Ioffe2005} B. L. Ioffe, Prog. Part. Nucl. Phys. {\bf 56} (2006) 232.

\bibitem{NarisonBook} S. Narison, Camb. Monogr. Part. Phys. Nucl. Phys. Cosmol. {\bf 17} (2002) 1.

\bibitem{Kho9801} A. Khodjamirian and R. Ruckl, Adv. Ser. Direct. High Energy Phys. {\bf 15} (1998) 345.
\bibitem{PDG} C. Amsler et al, Phys. Lett. {\bf  B667} (2008) 1.

\bibitem{Close2002} F. E. Close and N. A. Tornqvist, J. Phys.  {\bf G28} (2002) R249.

\bibitem{ReviewScalar}   C. Amsler and N. A. Tornqvist, Phys. Rept. {\bf 389} (2004) 61.

\bibitem{Wang0708} Z. G. Wang, Chin. Phys. {\bf C32} (2008) 797.

\bibitem{ChSh3} W. Lucha,  D. Melikhov and S. Simula,   Phys. Rev. {\bf D76} (2007) 036002.

\bibitem{LuchaPLB} W. Lucha, D. Melikhov and S. Simula, Phys. Lett. {\bf B687} (2010) 48.

\bibitem{Ebert05} D. Ebert,  R. N. Faustov and V. O. Galkin,  Phys. Lett. {\bf B634} (2006) 214.

\bibitem{Ebert0808} D. Ebert,  R. N. Faustov and V. O. Galkin,  Eur. Phys. J. {\bf C58} (2008) 399.


\bibitem{Maiani2004} L. Maiani, F. Piccinini, A. D. Polosa and V. Riquer, Phys. Rev. Lett. {\bf 93} (2004) 212002.

\bibitem{Maiani20042} L. Maiani, F. Piccinini, A. D. Polosa and V. Riquer, Phys. Rev. {\bf D71} (2005) 014028.


\bibitem{Polosa0902}  N. V. Drenska, R. Faccini and  A. D. Polosa, Phys. Rev. {\bf D79} (2009) 077502.

\bibitem{Ali-2010} A. Ali, C. Hambrock, I. Ahmed and M. Jamil Aslam, Phys. Lett. {\bf B684} (2010) 28.


\bibitem{Matheus0608} R. D. Matheus, S. Narison,  M. Nielsen and J.M. Richard,   Phys. Rev. {\bf D75} (2007)
014005.


\end{thebibliography}
\end{document}